\documentclass[11pt,letterpaper,reqno]{amsart}
\usepackage[utf8]{inputenc}
\usepackage[T1]{fontenc}
\usepackage{lmodern}
\usepackage{newunicodechar}
\newunicodechar{，}{,}
\usepackage[letterpaper,margin=1.05in]{geometry}

\usepackage{amsmath,amssymb,amsthm,mathtools}
\usepackage{microtype}
\usepackage{xcolor}
\usepackage{tikz}
\usepackage{longtable,array}
\usetikzlibrary{arrows.meta,calc,decorations.pathreplacing}

\definecolor{RevisionGreen}{RGB}{0,110,40}

\usepackage[colorlinks=true,linkcolor=blue,citecolor=blue,urlcolor=blue]{hyperref}
\hypersetup{pdftitle={Cantor spectrum at a fixed Diophantine frequency},
 pdfauthor={Rui Han, Wilhelm Schlag, Fan Yang},
 pdfsubject={Regular segments, resonances, localized eigenvalues, and regularity of the density of states}}
 
\numberwithin{equation}{section}
\newtheorem{theorem}{Theorem}[section]
\newtheorem{lemma}[theorem]{Lemma}
\newtheorem{proposition}[theorem]{Proposition}
\newtheorem{prop}[theorem]{Proposition}
\newtheorem{corollary}[theorem]{Corollary}

\theoremstyle{definition}
\newtheorem{definition}[theorem]{Definition}
\theoremstyle{remark}

\newcommand{\T}{\mathbb T}
\newcommand{\C}{\mathbb C}
\newcommand{\R}{\mathbb R}
\newcommand{\Z}{\mathbb Z}

\def\CZ{{\mathcal{C}_Z}}

\DeclareMathOperator{\mes}{mes}

\DeclareMathOperator{\dist}{dist}
\DeclareMathOperator{\spec}{spec}

\allowdisplaybreaks[1]
\usepackage{xcolor}

\title[Cantor spectrum at a fixed frequency]{Cantor spectrum at a fixed Diophantine frequency}

\author[R.\ Han]{Rui Han}
\address{Department of Mathematics \\ Louisiana State University  \\  Baton Rouge, LA 70803, USA}
\email{rhan@lsu.edu}

\author[W.\ Schlag]{Wilhelm Schlag}
\address{Department of Mathematics \\ Yale University \\ New Haven, CT 06511, USA}
\email{wilhelm.schlag@yale.edu}

\author[F.\ Yang]{Fan Yang}
\address{Department of Mathematics \\ Louisiana State University  \\  Baton Rouge, LA 70803, USA}
\email{yangf@lsu.edu}

\begin{document}

\begin{abstract}
We establish that the spectrum of one-frequency analytic quasiperiodic Schr\"odinger operators exhibits Cantor structure in the regime of positive Lyapunov exponents for every fixed Diophantine frequency.
The proof follows the general strategy of the gap construction by Goldstein and the second author, but avoids the elimination of double and triple resonances used in the original argument.
A novel ingredient is the use of H\"older continuity of the integrated density of states (IDS) to exclude energies at which eigenvalue parametrizations in finite volume have small slopes.
We also use a linear bound in the volume on the number of zeros of Dirichlet determinants, and a pigeonholing argument to obtain lower bounds on the distances between eigenvalues.
The crossing of two translated Rellich functions with opposite slopes produces a local maximum whose eigenfunctions have exponentially small boundary values.
Together with finite volume eigenvalue separation, this forces the DOS mass to concentrate near the maximum.
However, on an interval of constancy of the acceleration, the Lyapunov exponent is analytic, and hence the DOS is Lipschitz there.
This ultimately leads to a contradiction near the maximum.
\end{abstract}

\thanks{The first author is partially supported by NSF DMS-2143369. The second author is partially supported by the United States NSF through grant DMS-2350356. The second author thanks LSU for its hospitality. }

\maketitle

\setcounter{tocdepth}{1}
\tableofcontents

\section{Introduction}\label{sec:intro}

We consider the quasi-periodic Schr\"odinger operators
\begin{align*}
 [H(x)u](j)=-u(j-1)-u(j+1)+V(x+j\omega)u(j),
\end{align*}
on $\ell^2(\Z)$ where $V:\T\to\R$ is real analytic and $\T:=\R/\Z$. 
The frequency is fixed and satisfies
\begin{align}\label{eq:DC}
 \|k\omega\|_{\T}\geq c_\omega|k|^{-\tau_\omega}, \qquad k\in\Z\setminus\{0\},
\end{align}
for some $c_\omega>0$ and finite $\tau_\omega>1$. The spectrum $\Sigma_\omega$ is independent of the phase $x$.
Let $L(y,E)$ be the Lyapunov exponent at complex phase $x+iy$, and $\kappa(E)$ be the acceleration, whose definitions are recalled below.

\begin{theorem}\label{thm:cantor}
Let $V$ be real analytic and let $\omega$ satisfy \eqref{eq:DC}. On the open set
\begin{align*}
 \mathcal O_+=\{E\in\R:L(0,E)>0\},
\end{align*}
$\Sigma_\omega\cap\mathcal O_+$ is relatively closed in $\mathcal O_+$, has no isolated points, and has empty interior. If
$L(0,E)>0$ for every $E\in\Sigma_\omega$, then $\Sigma_\omega$ is a Cantor set.
\end{theorem}

The question whether the spectrum of the almost Mathieu operator, with potential $V(x)=2\lambda\cos(2\pi x)$, $\lambda\neq 0$, is a Cantor set for every irrational frequency was known as the Ten Martini Problem.
Bellissard and Simon~\cite{BS82} proved Cantor spectrum for a dense $G_\delta$ set of $(\lambda,\omega)$ pairs.
Last~\cite{Last94} proved that the spectrum has zero Lebesgue measure, and hence is a Cantor set, at the critical coupling $|\lambda|=1$ for almost every frequency.
Puig~\cite{Puig04} proved Cantor spectrum for Diophantine frequencies and $|\lambda|\neq 1$, using localization, Aubry duality, and reducibility.
His duality-based gap-opening argument has been influential in subsequent developments.
Avila and Krikorian~\cite{AK06} established zero Lebesgue measure of the critical spectrum for every irrational frequency.
The full problem was settled by Avila and Jitomirskaya in~\cite{AJ09}.

For general real analytic potentials, Goldstein and the second author~\cite[Theorem~1.1]{GS11} proved Cantor spectrum in the positive Lyapunov exponent (or supercritical) regime for Diophantine frequencies outside an exceptional set of Hausdorff dimension zero.
Their result assumes a stronger Diophantine condition.
The proof is based on a multiscale analysis of resonances, working directly with resonances in finite volume and the associated Rellich functions and their associated eigenfunctions. 
Damanik, Goldstein, the second author, and Voda~\cite{DGSV} subsequently established homogeneity of the spectrum and quantitative gap estimates in the positive Lyapunov exponent regime for Diophantine frequencies.

More recently, Ge, Jitomirskaya, and You~\cite{GJY} built on Puig's duality and reducibility argument and proved Cantor spectrum for Type I operators at every irrational frequency.
For the supercritical acceleration one operators, Li, Xu, and Zhou~\cite{LXZ} proved all gaps predicted by the gap-labeling condition are open. 
Ge, Jitomirskaya, and You~\cite{GJYsuper} subsequently proved the full Cantor spectrum result for all  supercritical operators and irrational frequencies. 
Our main result provides an alternative approach to that result without any duality and reducibility. Our proof follows the multiscale approach of~\cite{GS11}, but keeps the Diophantine frequency fixed.

In the subcritical regime, gap-opening arguments often rely on reducibility or almost reducibility of the Schr\"odinger cocycle. 
This approach goes back to the perturbative works of Moser--P\"oschel~\cite{MP84} and Eliasson~\cite{E92}, and was developed nonperturbatively in~\cite{Puig06,AJ10}.
See also~\cite{AvilaARC,HS_arc} for more recent developments in almost reducibility.

Cantor spectrum also occurs in other physically motivated magnetic models, including the extended Harper's model~\cite{HanDry}, Graphene model~\cite{BHJ19}, and Lieb lattice model~\cite{GSYLieb}. 
However, in the unitary setting, different spectral phenomena may occur. 
For example, certain electric quantum walks have the entire unit circle as their spectrum~\cite{Yang_unit}.

In sharp contrast to the one-frequency case, Goldstein, Schlag, and Voda~\cite{GSV} showed that multi-frequency quasi-periodic Schr\"odinger operators can have interval spectrum in the positive Lyapunov exponent regime.

\subsection{Outline of the proof}
\ 

This work began as an attempt to create a study guide for the delicate Cantor spectrum argument in~\cite{GS11}. 
Our original plan was to combine the construction of complex zeros in that paper with the first two authors' global zero count for Dirichlet determinants via acceleration~\cite{HS1}.
We sought to show that arbitrarily close to each spectral energy $E\in\mathcal O_+$ with $\kappa(E)=K\geq1$, there is an energy $E'\in\mathcal O_+$ with $\kappa(E')\leq K-1$.
Iterating this reduction would yield a nearby energy of zero acceleration and hence a point outside the spectrum. 
This idea was motivated by the complex-zero-count mechanism in~\cite{GS11}, which we wanted to express in terms of acceleration.

As we developed this approach, we found that we could keep the frequency fixed by using H\"older continuity of the integrated density of states (IDS) to exclude energies at which eigenvalue parametrizations in finite volume have small slopes. 
This avoids the frequency elimination associated with double resonances in~\cite{GS11}, which was used there to enforce strong localization properties.
We also avoid the elimination of triple resonant frequencies by controlling the number of separated singular regions through a linear bound in the volume on the number of zeros. 
A pigeonholing argument then allows us to obtain sufficiently strong eigenvalue separation at a larger scale.
At this stage, our proof still proceeded through the construction of additional complex zeros and the aforementioned decrease of the acceleration.

Finally, we realized that on a hypothetical spectral interval in $\mathcal O_+$, the Lyapunov exponent is real analytic near an energy of minimal acceleration, and hence the DOS has a real analytic density there.
The DOS mass produced by the eigenfunctions near the local maximum constructed below already contradicts the resulting Lipschitz bound for the IDS.
Thus, the construction of additional complex zeros and the reduction argument for the acceleration are no longer needed. 
Our earlier construction of complex zeros may be of independent interest, and we intend to present that approach elsewhere.

\smallskip

Next, we present the outline of the simplified proof.
Suppose that an interval of positive Lyapunov exponents lies in the spectrum. 
We pass to a smaller interval on which the acceleration is constant, denoted by $\kappa=K\geq1$, by choosing a minimal value of the acceleration. 
By \cite[Proposition~5]{Avila}, the Lyapunov exponent is real analytic on this interval.
The Thouless formula implies that the DOS has a real analytic density there. In particular, the IDS is locally Lipschitz.

Next we follow the basic construction from~\cite{GS11}.
Dirichlet eigenvalue parameterizations in finite volume\footnote{Following common practice, cf.~\cite{GS08}, we refer to these as \emph{Rellich functions}.} of opposite slopes at a smaller scale are made to cross via a translation by a multiple of~$\omega$.
This produces a local maximum of a Rellich function over a larger interval on~$\Z$. 
The goal is to show that such a maximum is attained at a sufficiently isolated eigenvalue with sufficiently small boundary values of its eigenfunction, which leads to a contradiction with the Lipschitz continuity of the IDS.

In order to construct two regular Rellich functions with appropriate slopes at the smaller scale, which we  call the two {\it parent curves}, we first use the H\"older continuity of the IDS to discard energies where Rellich functions could have small slopes.
Combining the Sard theorem estimates of~\cite[Section~11]{GS08} and \cite[Proposition~10.11]{GS11}, which we reprove here in Section~\ref{sec:orientation}, with a Borel--Cantelli argument, we obtain a set of energies of full measure with respect to the DOS, for which finite-volume eigenvalue branches have quantitatively large slopes.

In Section~\ref{sec:funf} we choose a generalized eigenvalue from this set via Schnol's theorem and construct the opposite spectral segments.
The generalized eigenfunction supplies many real crossings.
Zero counts for shorter endpoint determinants allow us to disregard crossings corresponding to eigenfunctions with significant mass near the edges.
Thus, we obtain two regular segments with opposite slopes and a common energy image, as in~\cite[Proposition~12.15]{GS11}.

To construct a local maximum with the required spectral control, we need two larger spatial scales.
The zero count in Lemma~\ref{lem:count-bounds} permits at most $\CZ$ separated singular regions within a polynomial spatial range.
Pigeonholing among $\CZ+1$ spatial annuli allows us to choose an interval $\Lambda_{\rm res}$ of length $q$ containing both parent Dirichlet intervals, and a larger interval $\Lambda\supset\Lambda_{\rm res}$ of length $Q\sim Dq$ such that $\Lambda\setminus\Lambda_{\rm res}$ has no singular sites.
Section~\ref{sec:localization} and Proposition~\ref{prop:FSWgap} then produce endpoint errors at most $2e^{-cQ}$ and eigenvalue separation at least $e^{-C_5q}$.
The latter spectral gap argument is due to Fr\"ohlich, Spencer, and Wittwer~\cite{FSW90}.
This pigeonhole argument replaces the frequency elimination of triple resonances in~\cite[Section~14]{GS11}.

Section~\ref{sec:local-max} translates the opposite segments to form a crossing and obtains a local maximum of a Rellich function of $H_\Lambda$, following \cite[Lemmas~15.1(i), 15.4, and Corollary~15.5]{GS11}.
At this maximum the first derivative vanishes, while the spectral gap lower bound $g=e^{-C_5q}$ bounds the second derivative by $C/g$.
Hence the eigenvalue varies by at most $Cr^2/g$ on a phase interval of length $2r$, where $r=e^{-(C_5+1)q}$.
Together with the endpoint errors $2e^{-cQ}$, the spectral theorem then implies DOS has mass at least $3r/(2Q)$ in an energy interval of length $C(r^2/g+e^{-cQ})$.
Choosing $D$ sufficiently large, this contradicts the Lipschitz bound.

Although we have highlighted the similarities with~\cite{GS08,GS11}, this paper can be largely read independently of these references.
Unlike~\cite{GS11}, which is constructive, our proof proceeds by contradiction.
In particular, we do not construct pre-gaps, and no comparison of zero counts at different scales is needed.

\subsection{AI Declaration}
\ 

ChatGPT was used to help us simplify the Cantor spectrum argument in~\cite{GS11}, initially with the same frequency exclusions. For our original approach, it suggested choosing an energy of minimal acceleration to avoid the need for an induction.

ChatGPT also helped optimize the numerous scales appearing in~\cite{GS11}, allowing us to work with polynomial scales $n^{-p}$ instead of sub-exponentially small scales. 
This is a technical advantage but not essential for the proof of the theorem.
ChatGPT simplified the proof of Lemma~\ref{lem:pollocmax}, which is based on \cite[Lemma~15.4]{GS11}, and the slopes Lemma~\ref{lem:kleinsten} which compresses the Sard's theorem arguments from~\cite{GS11}.

These simplifications led to an earlier manuscript that still involved the creation of many complex zeros and the subsequent decrease of the acceleration.
We then further simplified and revised the argument using the aforementioned analyticity of the Lyapunov exponent and of the DOS, arriving at the shorter version presented here.

\section{Preliminaries}
\label{sec:basictools}

We fix the frequency $\omega$ throughout and assume that it satisfies~\eqref{eq:DC}. 
All constants and thresholds below may depend on $\omega$ and $V$ and we do not indicate this dependence. 
We suppress~$\omega$ from most notations, but denote the spectrum of the operator by $\Sigma_\omega$, and the IDS and the DOS measure by $\mathcal{N}_{\omega}$ and $\nu_{\omega}=d\mathcal{N}_{\omega}$, respectively. We write $e(\zeta)=e^{2\pi i\zeta}$ and let 
\begin{align*}
 \T_r=\{\zeta\in\C/\Z:|\operatorname{Im}\zeta|<r\}
\end{align*}
be the periodic strip of width $2r$. We denote the open disk of center $z$ and radius $r$ by  $\mathcal D(z,r)$.
For a holomorphic function $g$ on a domain $U\subset\C/\Z$, let $Z_g(U)$ denote the number of zeros of $g$ in $U$, counted with multiplicity.
From now on, we fix an arbitrary compact energy interval $\mathcal J\Subset\mathcal O_+$. The constants for the rest of the paper will be chosen uniformly in $E\in \mathcal{J}$.

\subsection{Finite volume truncations}\label{subsec:finite}
\

For a finite interval $\Lambda\subset \Z$, let $H_\Lambda$ be the  restriction of $H$ on $\Lambda$ with Dirichlet boundary conditions. For complex phase $\zeta=x+iy$, set 
\begin{align*}
 F_\Lambda(\zeta,E)=\det(H_\Lambda(\zeta)-E),\qquad G_\Lambda(\zeta,E)=(H_\Lambda(\zeta)-E)^{-1},
\end{align*}
whenever the inverse exists.  

We write $E_j^\Lambda(x)$, the \emph{Rellich functions} of $H_\Lambda$, for its ordered and continuous eigenvalue parametrizations and let $\psi_j^\Lambda(x)$ be the associated $\ell^2$ normalized eigenfunctions (real for $x\in\R$ and unique up to a sign). 
Since the eigenvalues of $H_\Lambda(x)$ are simple, the $E_j^\Lambda$ are real analytic in $x$ by Rellich's theorem \cite[Ch.~II, \S6]{Kato}.

Depending on the compact energy interval $\mathcal{J}$, we choose a strip~$\T_\varrho$ with $\varrho>0$ sufficiently small so that $V$ is holomorphic and the determinant estimates in Section~\ref{sec:LDT} hold in a fixed neighborhood of $\overline{\T_\varrho}$.
The constant $M_V:=\max(1,\sup_{\zeta\in \overline{\T_{\varrho}}} |V'(\zeta)|)$ will be used throughout.  By first order eigenvalue perturbations 
$|\partial_xE_j^\Lambda(x)|\leq M_V$.
Moreover, for any $\zeta_1, \zeta_2\in \overline{\T_{\varrho}}$,
\begin{align*}
 \|H_\Lambda(\zeta_1)-H_\Lambda(\zeta_2)\| \leq M_V\|\zeta_1-\zeta_2\|_{\C/\Z}
\end{align*}

\subsection{Transfer matrices, Lyapunov exponents and Acceleration}
\

As usual,  monodromy matrices are defined as follows
\begin{align*}
 M_{[\ell,m]}(\zeta,E)=  
 \begin{pmatrix}
 V(\zeta+m\omega)-E&-1\\
 1&0
 \end{pmatrix} \begin{pmatrix}
 V(\zeta+(m-1)\omega)-E&-1\\
 1&0
 \end{pmatrix} \cdots \begin{pmatrix}
 V(\zeta+\ell\omega)-E&-1\\
 1&0
 \end{pmatrix}.
\end{align*}
For $\ell<m$, we have the well-known identity via determinants
\begin{align}\label{eq:M=ffff}
 M_{[\ell,m]}(\zeta,E)= \begin{pmatrix}
 F_{[\ell,m]}(\zeta,E)&-F_{[\ell+1,m]}(\zeta,E)\\
 F_{[\ell,m-1]}(\zeta,E)&-F_{[\ell+1,m-1]}(\zeta,E)
 \end{pmatrix}.
\end{align}
The Lyapunov exponent is defined as
\begin{align*}
L(y,E)=\lim_{n\to\infty}L_n(y,E)=\lim_{n\to\infty}\frac1n\int_0^1\log\|M_n(x+iy,E)\|\,dx, 
\end{align*} with $M_n=M_{[1,n]}$.
We abbreviate $L_n(E)=L_n(0,E)$ and $L(E)=L(0,E)$.  $L(E)$ is non-negative since $\det M_n(\zeta,E)=1$. 
We also have $L_n(y,E)\geq L(y,E)$ by sub-additivity. \cite{GS01}~established that $L(E)$ is H\"older continuous in $E$ for strong Diophantine frequencies, and it was later shown to be continuous in $E$ for any irrational $\omega$, see \cite{BJ}.  Set
\begin{align}\label{eq:positive-LE}
 L_*:=\min_{E\in\mathcal J}L(E)>0.
\end{align} 
For real $E$, $L(y,E)$ is even and convex in $y$.  Recall the acceleration introduced by Avila in \cite{Avila}: 
\begin{align}\label{def:acc}
 \kappa(E)=\frac{1}{2\pi}\partial_y^+L(0,E)\geq 0.
\end{align}
Avila proved the quantization of the acceleration, see~\cite[Theorem~5]{Avila}, 
\begin{align*}
 \frac{1}{2\pi}\partial_y^+L(y,E)\in\Z. 
\end{align*}
This plays a pivotal role in his global theory. Moreover, by~\cite[Proposition~5]{Avila}, the restriction of $L(E)$ to the interior of $\{E\in\mathcal O_+:\kappa(E)=K\}$, is real analytic. 
This means that it admits a holomorphic extension to a complex neighborhood of that set.

\subsection{Large deviation estimates and nearby zeros}\label{sec:LDT}
\

We rely on the large deviation estimates from \cite[Lemma 2.1]{BG00} and \cite[Proposition 2.11]{GS08}. These estimates apply to any $\omega$ satisfying~\eqref{eq:DC}.
To be precise, we may choose $0<\theta<1/2$ and $C_1\geq 1$ so that, for large enough $n$ and all intervals $|\Lambda|=n$, uniformly in $E\in\mathcal J$ and $|y|\leq\varrho$,
\begin{align}
 &\log|F_\Lambda(x+iy,E)|
 \le nL_n(y,E)+C_1n^{1-\theta},
 \label{eq:wtoben}\\
&\mes\left\{x:
\log|F_\Lambda(x+iy,E)|
<nL_n(y,E)-n^{1-\theta}
\right\}
\le e^{-n^\theta}.
\label{eq:wtldt}
\end{align}
The Lyapunov exponent is Lipschitz continuous in the $y$ direction, see for example~\cite[Corollary 4.2]{GS08}. We fix $C_2>0$ such that 
\begin{align}\label{eq:Lip}
 |L_n(y,E)-L_n(y',E)|\leq C_2|y-y'|,
\end{align}
for all $E\in\mathcal J$ and $|y|,|y'|\leq\varrho$, uniformly in $n$.

Let $a:=1-\theta/2$. The following consequence of Cartan's estimate and the maximum principle converts  large deviations of the determinant  into  nearby  zeros in the phase variable.  This is a standard tool, see~\cite[Lemma~2.2]{GS08} and~\cite[Lemma~2.18]{GS11}. We present its proof for the sake of completeness.

\begin{lemma}\label{lem:Cartan}
There are constants $C_3\geq 1$ and $c_4>0$ such that the following properties hold for every sufficiently large $n$, every interval $\Lambda\subset\Z$ with $|\Lambda|=n$, and every $E\in\mathcal J$. Let $\zeta_0=x_0+iy_0$ be such that $\mathcal D(\zeta_0,8/n)\subset\T_\varrho$. If $D\geq C_3n^{1-\theta}$ and
\begin{align*}
 \log|F_\Lambda(\zeta_0,E)|
 \leq nL_n(y_0,E)-D,
\end{align*}
then $F_\Lambda(\cdot,E)$ has a zero $\zeta$ with
\begin{align}\label{eq:wtzerd}
 \|\zeta-\zeta_0\|_{\C/\Z}
 \leq \frac{1}{n}
 \exp\left(-\frac{c_4D}{n^{1-\theta}}\right).
\end{align}
In particular, for $x\in\T$, the condition
\begin{align}\label{eq:deepSL}
 \log|F_\Lambda(x,E)|<nL_n(E)-C_3n^a
\end{align}
implies that $F_\Lambda(\cdot,E)$ has a zero $\zeta$ satisfying
\begin{align}\label{eq:tiefsl0}
 \|\zeta-x\|_{\C/\Z}\leq\rho_n,
 \qquad \rho_n:=\frac1n e^{-n^{\theta/2}}.
\end{align}
\end{lemma}

\begin{proof}
Let $R=\frac{7}{n}$ and $h:=nL_n(y_0,E)$. By \eqref{eq:wtldt}, there exists a {\em good point} $\zeta_g:=x_g+iy_0\in\T_\varrho$ not in the exceptional set such that
\begin{align*}
 \|\zeta_g-\zeta_0\|_\T<\frac{1}{n}, \qquad
 \log|F_\Lambda(\zeta_g,E)|
 \geq h-n^{1-\theta}.
\end{align*}
Thus $\zeta_0\in\mathcal D(\zeta_g,\frac{1}{n})$ and $\mathcal D(\zeta_g,R)\subset  \mathcal D(\zeta_0,\frac{8}{n})\subset\T_\varrho$.
For $\zeta=x+iy\in\mathcal D(\zeta_g,R)$, by \eqref{eq:wtoben}, \eqref{eq:Lip}, we have
\begin{align*}
 \sup_{\mathcal D(\zeta_g,R)}
 \log|F_\Lambda(\zeta,E)|
 &\leq h+C_1n^{1-\theta}+7C_2\\
 &\leq h+(C_1+1)n^{1-\theta}.
\end{align*}
The difference between this upper bound on $\mathcal{D}(\zeta_g,R)$ and the good-point lower bound at $\zeta_g$ is at most $(C_1+2) n^{1-\theta}$.
By Cartan's estimate, there is an absolute constant $C_*$ such that one can remove an exceptional set $\mathcal X$, a finite union of open disks with total radii at most
\begin{align*}
 C_*R\exp\left(
 -\frac{D}{2C_*(C_1+2)n^{1-\theta}}\right),
\end{align*}
and
\begin{align*}
 \log|F_\Lambda(\zeta,E)|\geq h-\frac D2,
 \qquad
 \zeta\in\mathcal D(\zeta_g,R/6)\setminus\mathcal X.
\end{align*}
Since $\log|F_\Lambda(\zeta_0,E)|\leq h-D$, we have $\zeta_0\in\mathcal X$.

Let $\mathcal X_0$ be the connected component of $\mathcal X$ that contains $\zeta_0$. 
Increasing $C_3$ if necessary, its diameter is less than $1/(6n)$, so $\overline{\mathcal X_0}\subset \mathcal D(\zeta_g,R/6)$, and
\begin{align}\label{eq:minPhi0}
 \min_{\partial\mathcal X_0}
 \log|F_\Lambda(\cdot,E)|\ge h-\frac{D}{2}.
\end{align}
If $F_\Lambda(\cdot,E)$ does not vanish in $\mathcal X_0$, then $\log|F_\Lambda(\cdot,E)|$ is harmonic there. By the minimum principle and~\eqref{eq:minPhi0},
\begin{align*}
\log|F_\Lambda(\zeta_0,E)| \geq \min_{\partial\mathcal X_0} \log|F_\Lambda(\cdot,E)| \geq h-\frac{D}{2},
\end{align*}
a contradiction. 
Hence $F_\Lambda(\cdot,E)$ has a zero in $\mathcal X_0$ and 
\begin{align*}
 \|\zeta-\zeta_0\|_{\C/\Z}
 &\leq\operatorname{diam}\mathcal X_0\\
 &\leq\frac{14C_*}{n} \exp\Big(-\frac{D}{4C_*(C_1+2)n^{1-\theta}}\Big)\\
 &\leq\frac1n \exp\left(-\frac{c_4D}{n^{1-\theta}}\right),
\end{align*}
with $c_4:=1/(8C_*(C_1+2))$, after increasing $C_3$ if necessary. This proves~\eqref{eq:wtzerd}. Finally, increasing $C_3$ further  to ensure that  $C_3\geq 1/c_4$ in \eqref{eq:wtzerd} yields~\eqref{eq:tiefsl0}.
\end{proof}

We use the threshold in \eqref{eq:deepSL} to define a resonant interval.
\begin{definition} 
Let $(x,E)\in\T\times\R$. An interval $\Lambda\subset \Z$ with $|\Lambda|=n$ is called {\it non-resonant} for $(x,E)$ if
\begin{align}\label{eq:fcgood}
\log|F_{\Lambda}(x,E)| \geq nL_n(E)-C_3n^a,
\end{align}
and {\it resonant} otherwise.
\end{definition}

\subsection{Total zero count of Dirichlet determinants}
\

From now on we fix $\varrho_0$ such that  $0<\varrho_0<\min\{\varrho/8,1/16\}$.
We record the following elementary upper bound for the total zero count of $F_{\Lambda}$ in the fixed strip $\T_{\varrho_0}$. 
A sharper asymptotics of zero counts in terms of the acceleration, including extensions to block Jacobi operators, can be found in~\cite{HS1,HSJacobiLoc,HSBlockHolder}. Such sharp bounds are not needed here.

\begin{lemma}\label{lem:count-bounds}
There exists an integer $\CZ \geq1$ such that for all large $n$ every interval $\Lambda$ with $|\Lambda|=n$, and every $E\in\mathcal J$, the total zero count of $F_{\Lambda}(\cdot,E)$  obeys
\begin{align}\label{eq:globbud}
 Z_{F_\Lambda(\cdot,E)}(\T_{\varrho_0})\leq \CZ\,  n.
\end{align}
\end{lemma}

\begin{proof}
Fix $0<\varrho_0<\min\{\varrho/8,1/16\}$ and cover $\T_{\varrho_0}$ by finitely many disks $\mathcal D(c_j,2\varrho_0)$, with $c_j\in\T$. By~\eqref{eq:wtldt}, for large $n$ there is a real point $x_j$ with $\|x_j-c_j\|_\T<\varrho_0/4$ such that
\begin{align*}
 \log|F_\Lambda(x_j,E)|\geq nL_n(E)-n^{1-\theta}\geq-n.
\end{align*}
Boundedness of $V$ on the fixed strip and of $E\in\mathcal J$ implies that 
\begin{align*}
 \sup_{\mathcal D(x_j,4\varrho_0)}\log|F_\Lambda(\cdot,E)|\leq Cn.
\end{align*}
Jensen's formula therefore yields
\begin{align*}
 Z_{F_\Lambda(\cdot,E)}(\mathcal D(x_j,3\varrho_0))\log(4/3)\leq(C+1)n.
\end{align*}
Since $\mathcal D(c_j,2\varrho_0)\subset\mathcal D(x_j,3\varrho_0)$, summing over the fixed cover proves~\eqref{eq:globbud}. 
\end{proof}

\subsection{Regularity of the density of states measure}
\

H\"older continuity of the IDS plays an important role in our proof. 
For Diophantine frequencies, see~\cite{GS01,GS08,HS1,HSBlockHolder}, and extensions to certain Liouville frequencies can be found in~\cite{HanZhang,HS_liou_holder}. The estimates of~\cite{HSBlockHolder} apply throughout $\mathcal O_+$ for every $\omega$ satisfying \eqref{eq:DC}.
Recall that we fixed $\mathcal J\Subset \mathcal{O}_+$. There exist $\beta>0$ and $C_{\mathcal J}^{(\beta)}>0$ such that
\begin{align}\label{eq:holder-intro}
 |\mathcal N_\omega(E_2)-\mathcal N_\omega(E_1)| \leq C_{\mathcal J}^{(\beta)} |E_2-E_1|^\beta,\qquad \text{for }E_1,E_2\in \mathcal J.
\end{align}
Equivalently, in terms of the DOS measure,
\begin{align}\label{eq:holderEin}
 \nu_\omega(\mathcal J')\leq C_{\mathcal J}^{(\beta)} |\mathcal J'|^\beta \qquad\text{for any interval } \mathcal J'\subset \mathcal J.
\end{align}
H\"older continuity plays a crucial role in the proof, it is used to exclude energies in the range of Rellich functions with small slopes.
It was shown in~\cite{HSBlockHolder} that the H\"older exponent can be any $0<\beta<1/(2\kappa(E))$ near each $E\in \mathcal J$. Here we do not need the sharp exponent, any $\beta>0$ would suffice.

The next lemma provides a hypothetically stronger regularity of the IDS, only applicable under the assumption that the spectrum contains an interval.
\begin{lemma}\label{lem:andos}
Let $I \subset\R$ be an open interval on which $E\mapsto L(E)$ is real analytic. Then $\nu_\omega$ has a real analytic density on $I$. In particular, for every compact interval $J\Subset I$, there exists $C_{J}>0$ such that
\begin{align}\label{eq:doslip}
 \nu_\omega(J')\leq C_J|J'|,\qquad J'\subset J.
\end{align}
\end{lemma}

\begin{proof}
By the Thouless formula, the Lyapunov exponent and the IDS are related through the Hilbert transform. 
By standard properties of the Hilbert transform, the analyticity of $L$ implies that $\nu_\omega$ has a real analytic density on $J$,  see also \cite[Section~10]{GS01}.
The Lipschitz bound \eqref{eq:doslip} is an immediate consequence.
\end{proof}

\subsection{Separation and mobility of eigenvalues}
\

The following separation estimate is a standard fact for finite Jacobi matrices, and we leave the proof to the reader.
\begin{lemma}\label{lem:jacobi-gap}
Let $J$ be a $q\times q$ real symmetric tridiagonal matrix, with off-diagonal entries $-1$ and diagonal entries of absolute value at most $M$, where $M\geq2$. Its eigenvalues are simple, and distinct eigenvalues are separated by at least $e^{-C(M)q}$, where $C(M)>0$ depends only on $M$.
\end{lemma}

The following sharper bound by Fr\"ohlich--Spencer--Wittwer states that if two orthogonal eigenfunctions both have most of their mass on some shorter interval $\Lambda_0\subset \Lambda$, then their eigenvalues are exponentially separated in terms of~$|\Lambda_0|$.

\begin{proposition}\label{prop:FSWgap}\cite[Lemma 4.1]{FSW90}
Let $\Lambda_0$ be a sub-interval of $\Lambda\subset \Z$ with $|\Lambda_0|\geq 2$. Let $\psi_1,\psi_2$ be two normalized eigenfunctions of $H_\Lambda(x)$ corresponding to distinct eigenvalues $E_1,E_2$. Suppose
\begin{align}\label{eq:masses34}
 \|\mathbf{1}_{\Lambda_0}\psi_j\|^2
 \geq\tfrac{3}{4},\qquad j=1,2.
\end{align}
Then
\begin{align}\label{eq:34gap}
 |E_1-E_2|\geq e^{-C_5|\Lambda_0|},
\end{align}
where $C_5=C(V)>0$ depends only on a uniform bound for the potential $V$.
\end{proposition}

The following lemma is a convenient and simple tool to extend analytic segments of Rellich graphs. 

\begin{lemma}\label{lem:transcom}
Let $\lambda:\T\to\R$ be real analytic, let $I=(E_-,E_+)$, and suppose that 
\begin{align*}
 \lambda(x)\in I\quad\Longrightarrow\quad|\lambda'(x)|>\tau>0.
\end{align*}
Every component of $\lambda^{-1}(I)$ is mapped monotonically and analytically onto $I$. If $\lambda^{-1}(I)\ne\varnothing$, components of both orientations occur. If also $|\lambda'|\leq M_V$, the phase length of each component is at least $|I|/M_V$.
\end{lemma}

\begin{proof}
On a component $(a,b)$ of $\lambda^{-1}(I$) the derivative has a fixed nonzero sign. 
Maximality implies $\lambda(a),\lambda(b)\in\{E_-,E_+\}$, and strict monotonicity makes $a,b$ distinct. 
Thus the image is $I$, and the inverse is analytic.  Periodicity implies equal numbers of positive and negative crossings of any level in $I$, so both orientations occur. 
Finally,
\begin{align*}
 |I|=|\lambda(b)-\lambda(a)|\leq M_V|b-a|,
\end{align*}
implies the phase length $|b-a|\geq |I|/M_V$.
\end{proof}

\section{Slopes of Rellich functions}
\label{sec:orientation}

In this section we study the derivatives (with respect to the phase) of finite-volume eigenvalues. 
By self-adjointness and simplicity of the eigenvalues of $H_{[1,n]}(x)$ and the Kato--Rellich theorem, the eigenvalue branches (Rellich functions) are analytic on~$\T$.

The main estimate is a Sard-type bound for the set of energies where some branch has small slope. The argument approximates $V$ by a trigonometric polynomial, uses the eigenvalue separation from Lemma~\ref{lem:jacobi-gap}, and then applies an algebraic counting step (via B\'ezout's theorem) to bound the number of monotonicity intervals. 
Related bounds appear in \cite[Proposition~10.11]{GS11}.  Unlike~\cite{GS11}, we keep all frequencies $\omega$ and use the elementary gap $g_n\ge e^{-Cn}$.

\begin{lemma}\label{lem:kleinsten}
For every sufficiently large $n$ and every $e^{-n^2/2}\leq\tau\leq1$, there exists a set
\begin{align*}
 \mathcal E_n(\tau) =\bigcup_{\ell=1}^{K_n}I_{n,\ell}, \qquad K_n\leq n^7,
\end{align*}
where the $I_{n,\ell}$ are closed intervals satisfying
\begin{align*}
 |I_{n,\ell}|\leq 4\tau,\qquad \mes\mathcal{E}_n(\tau)\leq 4n^7\tau,
\end{align*}
such that, for every $x\in\T$ and $1\leq j\leq n$, if 
\begin{align*}
 |\partial_xE_j^{[1,n]}(x)|\leq\tau \text{\ \ \ then\ \ \ }E_j^{[1,n]}(x)\in \mathcal{E}_n(\tau).
\end{align*}
The lower threshold for $n$ depends only on $V$.
\end{lemma}

\begin{proof}
Throughout the proof, constants denoted by $C$ depend only on $V$. For brevity, write
\begin{align*}
 H=H_{[1,n]}(x),\qquad E_j=E_j^{[1,n]}(x).
\end{align*}
By Lemma~\ref{lem:jacobi-gap} one has the uniform spectral gap 
\begin{align}\label{eq:trivgap}
 \min_{i\ne j}|E_i-E_j|\geq g_n:=e^{-Cn}.
\end{align}
By analyticity of $V$ there exists  a real trigonometric polynomial $V_n$ of degree $k_n\leq Cn^2$ such that
\begin{align*}
 \|V-V_n\|_{C^1(\T)}\leq\eta_n:=e^{-n^2}\ll \tau.
\end{align*}
Let $\widetilde H$ and $\widetilde E_j$ denote the corresponding matrix and its ordered eigenvalues. Then
\begin{align*}
 \|H-\widetilde H\|\leq\eta_n,\qquad \|\partial_xH-\partial_x\widetilde H\|\leq\eta_n.
\end{align*}
Let $\Pi_j$ and $\widetilde\Pi_j$ be the rank-$1$ orthogonal projections onto the eigenspaces
corresponding to $E_j$ and $\widetilde E_j$, respectively. For sufficiently large $n$, we conclude that 
$\eta_n\leq g_n/4$. Hence the positively oriented contour 
\begin{align*}
 \Gamma_j:=\{z\in\mathbb C:|z-E_j|=g_n/2\}
\end{align*}
encloses $E_j$ from the spectrum of $H$ and  $\widetilde E_j$ from the spectrum of
$\widetilde H$, but no other eigenvalues. Both projections therefore equal  Riesz projections the same
contour $\Gamma_j$. Integration of the resolvent identity over $\Gamma_j$ therefore yields
\begin{align}\label{eq:approxEn}
 |E_j-\widetilde E_j|\leq\eta_n,\qquad \|\Pi_j-\widetilde\Pi_j\|\leq C\eta_n/g_n.
\end{align}
By the Feynman--Hellmann formula 
\begin{align}
 |\partial_xE_j-\partial_x\widetilde E_j| &\leq \bigl|\operatorname{tr}((\Pi_j-\widetilde\Pi_j)\partial_xH)\bigr| +\bigl|\operatorname{tr} (\widetilde\Pi_j(\partial_xH-\partial_x\widetilde H))\bigr| \notag \\
 &\leq 2\|V'\|_{L^\infty(\T)}\|\Pi_j-\widetilde\Pi_j\| +\|\partial_xH-\partial_x\widetilde H\| \notag \\
 &\leq Ce^{-n^2+Cn} <e^{-n^2/2}\leq\tau\label{eq:approxSl}
\end{align}
for sufficiently large $n$, uniformly in $x$.
Cover $\T$ by the two closed semicircles parametrized by
\begin{align}\label{eq:2chart}
 e(x)=e(x_c)\frac{1+it}{1-it},\qquad x_c\in\{0,\tfrac12\},\quad -1\leq t\leq1, \qquad \frac{dx}{dt}=\frac1{\pi(1+t^2)}.
\end{align}
Here the derivative is taken using the lift $x=x_c+\pi^{-1}\arctan t$. On either chart, each potential term can be written as
\begin{align*}
 V_n(x+\ell\omega) =\frac{a_\ell(t)}{(1+t^2)^{k_n}}, \qquad a_\ell\in\mathbb R[t],\quad \deg a_\ell\leq2k_n.
\end{align*}
Consequently,
\begin{align*}
 P(t,E):=(1+t^2)^{nk_n}\det(\widetilde H-E)
\end{align*}
is a real polynomial in $(t,E)$ satisfying
\begin{align*}
 \deg_tP\leq2nk_n,\qquad \deg_EP=n,\qquad \deg_{t,E}P\leq2nk_n+n\leq Cn^3,
\end{align*}
where $\deg_{t,E}$ denotes the total degree.

The eigenvalues of $\widetilde H$ are simple, so $\partial_EP\ne0$ on every real eigenvalue graph. Implicit differentiation therefore implies
\begin{align*}
 \partial_x\widetilde E_j(x) =-\pi(1+t^2) \frac{\partial_tP(t,\widetilde E_j(x))} {\partial_EP(t,\widetilde E_j(x))}.
\end{align*}
Thus the condition $|\partial_x\widetilde E_j|\leq2\tau$ is equivalent, on the real eigenvalue graphs, to
\begin{align}
 P(t,E)=&0,\qquad B(t,E)\leq 0, \notag \\
 B(t,E):=&\pi^2(1+t^2)^2(\partial_tP)^2 -4\tau^2(\partial_EP)^2.\label{eq:sa-slope}
\end{align}
In particular,
\begin{align*}
 \deg_{t,E}B \leq2\deg_{t,E}P+2 \leq Cn^3.
\end{align*}
Let $U:=\gcd(P,B)$ and $P_0:=P/U$, with the gcd taken in $\mathbb R[t,E]$. No real eigenvalue point can
satisfy $U=P_0=0$, since at such a point
\begin{align*}
 \partial_EP=(\partial_EU)P_0+U(\partial_EP_0)=0,
\end{align*}
contrary to simplicity of eigenvalues of $\widetilde{H}$. By connectedness, each eigenvalue graph on the chart therefore lies entirely in either $\{U=0\}$ or $\{P_0=0\}$. On a graph contained in $\{U=0\}$, we have $B=0$ identically, since $U$ divides $B$. Hence the entire graph satisfies \eqref{eq:sa-slope}.

Our next step is to apply Bezout's theorem to $P_0,B$. In order to do so, we claim that $\gcd(P_0,B)=1$. If $B\equiv0$, then $P_0=1$. Otherwise, let $P_1$ be an irreducible factor common to $P$ and $B$, and let $m_1$ be its multiplicity in $P$ and $m_2$ be its multiplicity in $B$. It is enough to show that $m_2\geq m_1$.

If $m_1=1$, then $m_2\geq 1=m_1$. If $m_1\geq2$, then
\begin{align*}
 P_1^{m_1-1}\mid\partial_tP,\qquad P_1^{m_1-1}\mid\partial_EP,
\end{align*}
and hence
\begin{align*}
 P_1^{2m_1-2}\mid B.
\end{align*}
Hence $m_2\geq 2m_1-2\geq m_1$. Thus, in either case, $U$ contains $P_1$ with its full multiplicity in $P$, so $P_1\nmid P_0=P/U$. This proves the claim.

On the graphs contained in $\{P_0=0\}$, the zeros of $B$ belong to the algebraic intersection
\begin{align*}
 \{(t,E)\in\mathbb C^2:P_0(t,E)=B(t,E)=0\}.
\end{align*}
If $P_0$ is constant, the intersection is empty. Otherwise, since $P_0$ and $B$ are coprime, B\'ezout's theorem bounds the number of these points by
\begin{align*}
 (\deg_{t,E}P_0)(\deg_{t,E}B)\leq Cn^6.
\end{align*}
The zeros of $B$ partition the remaining eigenvalue graphs into intervals on which its sign is constant. Including possible isolated zeros, chart endpoints, and the complete graphs contained in $\{U=0\}$, we conclude that
\begin{align*}
 \{(t,E)\in[-1,1]\times\mathbb R: P(t,E)=0,\ B(t,E)\leq0\}
\end{align*}
has at most $Cn^6$ connected components.  Each component is a closed interval of a single Rellich graph, possibly reduced to a point. 
Its energy image is a closed interval of length at most $2\tau$, because $|\partial_x\widetilde E_j|\leq2\tau$ throughout the component and its phase length is at most one.

Define $\mathcal E_n(\tau)$ to be the union of the closed $\eta_n$-neighborhoods of these energy images, using both charts.  Adjusting $C$, there are at most $Cn^6$ such intervals, and each has length at most
\begin{align*}
 2\tau+2\eta_n\leq4\tau.
\end{align*}
Increasing the lower threshold for $n$, depending only on $V$, we have $Cn^6\leq n^7$. Therefore
\begin{align*}
 K_n\leq n^7,\qquad \mes\mathcal E_n(\tau)\leq4n^7\tau.
\end{align*}
Finally, if $|\partial_xE_j|\leq\tau$, then \eqref{eq:approxSl} implies $|\partial_x\widetilde E_j|\leq2\tau$.
Thus $\widetilde E_j$ belongs to one of the energy images constructed above, and \eqref{eq:approxEn} places $E_j$ in its closed $\eta_n$-neighborhood. 
\end{proof}

We now use \eqref{eq:holderEin} to exclude the  energies from flat sections of Rellich graphs at all sufficiently large scales, with a polynomial margin. 
The remaining set $\mathcal G$ has full DOS measure in $\mathcal J$.  In Section~\ref{sec:p_by_c} we pass from this statement to spectral measures.

\begin{proposition}\label{prop:fpfenster}
Recall that the IDS is $\beta$-H\"older continuous on the compact interval $\mathcal{J}\Subset \mathcal{O}_+$, see \eqref{eq:holderEin}. Choose
\begin{align*}
 p_2>p_1>9/\beta.
\end{align*}
There is a Borel set $\mathcal G\subset \operatorname{int}(\mathcal{J})$ with $\nu_\omega(\operatorname{int}(\mathcal{J})\setminus\mathcal G)=0$ such that, for every $E\in\mathcal G$ and all sufficiently large integers $n$,
\begin{align}\label{eq:disttoflach}
 \dist\bigl(E,\mathcal E_n(n^{-p_2})\bigr)>8n^{-p_1}.
\end{align}
In particular, for every interval $\Lambda$ with $|\Lambda|=n$,
\begin{align}\label{eq:buffer}
 |E_j^\Lambda(x)-E|\leq8n^{-p_1} \text{\ \ \ implies that\ \ \ }|\partial_xE_j^\Lambda(x)|>n^{-p_2}, \qquad x\in\T.
\end{align}
The lower threshold for $n$ may depend on $E$, but is independent of $\Lambda,x,j$.
\end{proposition}

\begin{proof}
Set the forbidden zone at scale $n$ as:
\begin{align}\label{def:Bn}
 \mathcal B_n=\mathcal E_n(n^{-p_2})+[-8n^{-p_1},8n^{-p_1}].
\end{align}
By Lemma \ref{lem:kleinsten} with $\tau=n^{-p_2}$, for large $n$, this set has a cover by at most $n^7$ intervals of length at most $4n^{-p_2}+16n^{-p_1}\leq 20n^{-p_1}$.
Hence
\begin{align}\label{eq:bad-dos}
 \nu_\omega(\operatorname{int}(\mathcal{J}) \cap\mathcal B_n) \leq C_{\mathcal J}^{(\beta)} n^7(20n^{-p_1})^\beta \leq Cn^{7-p_1\beta}\leq n^{-2}.
\end{align}
Here we used $p_1\beta>9$. Therefore,
\begin{align*}
 \sum_n\nu_\omega(\operatorname{int}(\mathcal{J}) \cap\mathcal B_n)<\infty.
\end{align*}
Borel--Cantelli yields $\mathcal G=\operatorname{int}(\mathcal{J})\setminus\limsup_n\mathcal B_n$ with the desired properties. 
If \eqref{eq:buffer} failed, Lemma~\ref{lem:kleinsten} and covariance would place an eigenvalue within $8n^{-p_1}$ of $E$ in $\mathcal E_n(n^{-p_2})$, contradicting \eqref{eq:disttoflach}.
\end{proof}

\section{Crossings and regular segments}
\label{sec:funf}

\subsection{Selecting a good generalized eigenvalue}\label{sec:p_by_c}
\

It is elementary that $\Sigma_\omega$ is closed and has no isolated points~\cite{Pastur80}.
Thus $\Sigma_\omega$ is perfect, and
$\Sigma_\omega\cap\mathcal O_+$ is relatively closed
in $\mathcal O_+$ and has no isolated points.
Since $\mathcal J\Subset\mathcal O_+$ was chosen
arbitrarily, it suffices to prove that
$\Sigma_\omega\cap\mathcal J$ has empty interior.
Suppose that $\Sigma_\omega\cap\mathcal{J}$ contains a nonempty open interval $I_{\mathrm{sp}}$. We now fix the parameters used in the rest of the proof. 
Choose $E_0\in I_{\mathrm{sp}}$ with
\begin{align*}
 \kappa(E_0)=K:=\min_{E'\in I_{\mathrm{sp}}}\kappa(E').
\end{align*}
The minimum exists because the acceleration is a finite nonnegative integer. It is clear that $K\geq1$. Indeed,
positive Lyapunov exponent and zero acceleration imply uniform hyperbolicity, and hence the energy would need to belong to the resolvent set by Johnson's theorem, cf.~\cite[Theorems~5 and 6]{Avila}. Upper semicontinuity and quantization imply that $\kappa(E)\leq K$ near
$E_0$, while minimality yields equality in a neighborhood of $E_0$ in $I_{\mathrm{sp}}$. 
Choose $\varepsilon_{\rm sp}>0$ so that $(E_0-2\varepsilon_{\rm sp},E_0+2\varepsilon_{\rm sp})\subset I_{\rm sp}$ is contained in this neighborhood with constant acceleration. 
By \cite[Proposition~5]{Avila}, $E\mapsto L(E)$ is analytic on this interval.
Let
\begin{align*}
 J_0:=[E_0-\varepsilon_{\rm sp}, E_0+\varepsilon_{\rm sp}].
\end{align*}
The Lipschitz bound of IDS therefore holds on $J_0$, see Lemma~\ref{lem:andos}.
Recall that the set $\mathcal G$ of good energies has full DOS measure in $J_0$.  Let $\mu_{x,j}$ be the spectral measure associated to the vector $\delta_j$ for $H(x)$, and let $\mu_x=\mu_{x,0}+\mu_{x,1}$.
Since $\delta_0,\delta_1$ form a cyclic set, $\operatorname{supp}\mu_x=\Sigma_\omega$. Hence $\mu_x(\operatorname{int}(J_0))>0$ for every $x$.
It is standard that
\begin{align*}
 \int_\T \mu_x(\operatorname{int}(J_0)\setminus\mathcal G)\,dx =2\nu_\omega(\operatorname{int}(J_0)\setminus\mathcal G)=0.
\end{align*}
Thus $\mu_x(\mathrm{int}(J_0)\cap\mathcal G)>0$ for almost every $x$.
By the Schnol--Berezanskii theorem~\cite{Far,HanShnol}, we can choose one such phase $x_\phi$ and an energy
$E_1\in\mathrm{int}(J_0)\cap\mathcal G$ whose normalized generalized
eigenfunction $\phi$ has at most linear growth:
\begin{align}\label{eq:gnorm}
 H(x_\phi)\phi=E_1\phi,\qquad
 \max\{|\phi(0)|,|\phi(-1)|\}=1,\qquad
 |\phi(j)|\leq C_\phi(1+|j|).
\end{align}
The finite-volume construction below applies to any generalized eigenvalue
$E_1\in\mathcal G\cap\mathrm{int}(\mathcal J)$.
We choose $E_1\in\mathrm{int}(J_0)$ so that the energy windows shrinking
to $E_1$ in Section~\ref{sec:dos-contradiction} eventually lie in $J_0$,
where the Lipschitz bound~\eqref{eq:doslip} applies.

\subsection{Real crossings from a generalized eigenfunction}\label{sec:cross}
\

The following lemmas will be used repeatedly. They relate zeros of $F_{\Lambda}(\cdot,E)$ near $x_0$ to resonances of the interval $\Lambda$ at $(x_0,E)$.

\begin{lemma}\label{lem:res_close}
If $\Lambda\subset \Z$ with $|\Lambda|=n$ sufficiently large is resonant at $(x,E)\in\T\times\mathcal J$, then there exists a Dirichlet eigenvalue $E_j^{\Lambda}(x)$ such that
\[
|E-E_j^{\Lambda}(x)|\leq M_V\rho_n.
\]
\end{lemma}
\begin{proof}
By Lemma~\ref{lem:Cartan}, a resonance of $\Lambda$ at $(x,E)$ yields a complex zero $\zeta$ of $F_{\Lambda}(\cdot,E)$ with $\|\zeta-x\|_{\C/\Z}\mathrel{\leq}\rho_n$.
Let $v$ be a normalized eigenvector satisfying $H_{\Lambda}(\zeta)v=Ev$. 
\begin{align*}
\dist\bigl(E,\sigma(H_{\Lambda}(x))\bigr)
&\leq \|(H_{\Lambda}(x)-E)v\| \\
&\leq \|H_{\Lambda}(x)-H_{\Lambda}(\zeta)\| \\
&\leq M_V\|x-\zeta\|_{\C/\Z}
\leq M_V\rho_n.
\end{align*}
Hence $\sigma(H_{\Lambda}(x))$ contains an eigenvalue $E_j^{\Lambda}(x)$ with $|E_j^{\Lambda}(x)-E|\leq M_V\rho_n$.
\end{proof}
\begin{lemma}\label{lem:no0nonres}
Let $E\in \mathcal{G}$ and suppose $\Lambda\subset\Z$ is an interval with $|\Lambda|=n=2\tilde{n}+1$ sufficiently large.
If $F_{\Lambda}(x,E)$ has no zeros in $x\in [x_0-4r,x_0+4r]$ for some $r$ with $\widehat{\rho}_n:=M_V n^{p_2}\rho_n\leq r\leq1/8$, then $\Lambda$ is non-resonant at $(x,E')$ for any $|x-x_0|\leq 2r$ and $|E'-E|\leq rn^{-p_2}$. In particular, if $\Lambda$ is resonant at $(x_0,E)$ then $F_{\Lambda}(x,E)$ vanishes somwhere on $[x_0-4r, x_0+4r]$. 
\end{lemma}
\begin{proof}
Assume for contradiction that $\Lambda$ is resonant at $(x,E')$ for some $|x-x_0|\leq 2r$ and $|E'-E|\leq rn^{-p_2}$.
By Lemma~\ref{lem:res_close}, there exists a Dirichlet eigenvalue $E_j^{\Lambda}(x)$ such that
\begin{align*}
|E_j^{\Lambda}(x)-E'|\leq M_V\rho_n.
\end{align*}
Therefore
\begin{align*}
|E_j^{\Lambda}(x)-E|\leq |E'-E|+|E_j^{\Lambda}(x)-E'|\leq rn^{-p_2}+M_V\rho_n<n^{-p_1}.
\end{align*}
Since $E\notin \mathcal B_n$, the Rellich functions have the uniform slope bound $|\partial_x E_j^{\Lambda}(\cdot)|\ge n^{-p_2}$ whenever the energy falls in $(E-4n^{-p_1},E+4n^{-p_1})$. 
Applying Lemma~\ref{lem:transcom}, the branch $E_j^{\Lambda}(\cdot)$ crosses $E$ at some $x_*$ with
\begin{align*}
|x_*-x|\leq r+M_V\rho_n n^{p_2}.
\end{align*}
In particular,
\begin{align*}
|x_*-x_0|\leq |x_*-x|+|x-x_0|\leq 3r+M_Vn^{p_2}\rho_n\leq 4r,
\end{align*}
so $F_{\Lambda}(x_*,E)=0$ occurs in $[x_0-4r,x_0+4r]$, contradicting the assumption that $F_{\Lambda}(\cdot,E)$ has no zeros there.
\end{proof}

This section produces many crossings using the generalized eigenfunction $\phi$ at energy $E_1$ and
phase~$x_\phi$, see~\eqref{eq:gnorm}. 
Fix $0<\delta<1/2$. For odd $n=2\tilde{n}+1$, define
\begin{align*}
 \Lambda_n=[-\tilde{n},\tilde{n}],\qquad \mathcal K_n=\{k\in\Z:|k|\leq (1-\delta)\tilde{n}\},\qquad |\mathcal K_n|=(1-\delta)n+O(1).
\end{align*}

\begin{lemma}\label{lem:gfLDT}
For every sufficiently large odd $n$ and every $k\in\mathcal K_n$,
the interval $\Lambda_n+k$ is resonant at $(x_\phi,E_1)$. Equivalently
\begin{align}\label{eq:odsub}
 \log|F_{\Lambda_n}(x_\phi+k\omega,E_1)|<nL_n(E_1)-C_3n^a.
\end{align}
\end{lemma}

\begin{proof}
Both $0$ and $-1$ have distance at least $\delta\tilde n-O(1)$ from the endpoints of
$[k-\tilde n,k+\tilde n]$. If \eqref{eq:odsub} failed, then
\begin{align*}
 |F_{[k-\tilde{n},k+\tilde{n}]}(x_\phi,E_1)|\geq\exp(nL_n(E_1)-C_3n^a).
\end{align*}
Cramer's rule and \eqref{eq:wtoben} imply, for $s=0,-1$,
\begin{align*}
 |G_{[k-\tilde{n},k+\tilde{n}]}(x_\phi,E_1;s,k-\tilde{n})| &\leq\exp\bigl(-(\delta\tilde n-O(1))L(E_1)+Cn^a\bigr), \\
 |G_{[k-\tilde{n},k+\tilde{n}]}(x_\phi,E_1;s,k+\tilde{n})| &\leq\exp\bigl(-(\delta\tilde n-O(1))L(E_1)+Cn^a\bigr).
\end{align*}
From the Poisson formula
\begin{align}\label{eq:Poisson}
 \phi(s)=G_{[k-\tilde{n},k+\tilde{n}]}(x_\phi,E_1;s,k-\tilde{n})\phi(k-\tilde{n}-1) +G_{[k-\tilde{n},k+\tilde{n}]}(x_\phi,E_1;s,k+\tilde{n})\phi(k+\tilde{n}+1),
\end{align}
and the linear growth of $\phi$, see \eqref{eq:gnorm}, one obtains
\begin{align*}
 1\leq|\phi(0)|+|\phi(-1)| \leq C_\phi n e^{-\delta L(E_1)n/4}<1
\end{align*}
for large $n$, a contradiction.
\end{proof}

\begin{corollary}\label{cor:genRZug}
For every sufficiently large odd $n$ and every $k\in\mathcal K_n$, there are an index $j(k)$ and a real phase
$x_{n,k}$ such that
\begin{align}\label{eq:grzug}
 E_{j(k)}^{\Lambda_n}(x_{n,k})=E_1, \qquad
 \|x_{n,k}-(x_\phi+k\omega)\|_\T\leq 4\widehat{\rho}_n, \qquad
 |\partial_xE_{j(k)}^{\Lambda_n}(x_{n,k})|>n^{-p_2}.
\end{align}
Thus, there are at least $(1-2\delta)n$ distinct real zeros of $F_{\Lambda_n}(\cdot,E_1)$.
\end{corollary}

\begin{proof}
Lemma~\ref{lem:gfLDT} implies that $\Lambda_n+k$ is resonant at $(x_\phi,E_1)$. 
By Lemma \ref{lem:no0nonres}, with $r=\widehat{\rho}_n$,  there is a real zero $x_{n,k}$ of $F_{\Lambda_n}(\cdot, E_1)$ in $[x_\phi+k\omega-4\widehat{\rho}_n, x_\phi+k\omega+4\widehat{\rho}_n]$.
For distinct $k,k'\in\mathcal K_n$,
$\|(k-k')\omega\|_\T\geq c_\omega n^{-\tau_\omega} \gg\widehat\rho_n$.
The crossings near $x_{\phi}+k\omega$ are therefore distinct. The lower bound on the slope follows from $E_1\notin \mathcal{B}_n$, namely $E_1$ is not in the forbidden zone at scale $n$.
\end{proof}

\subsection{Regular spectral segments with opposite slopes}
\label{sec:saturation}

\begin{definition}
Let $\Lambda\subset\mathbb Z$ be finite interval and let $H_\Lambda(x)$ be the Dirichlet restriction of $H(x)$. A
\emph{spectral segment} of $H_\Lambda$, or simply a {\em segment}, is a connected smooth curve
\begin{align*}
 \Gamma=\{(x,E(x)):\ x\in X\},
\end{align*}
where $X\subset\mathbb T$ is an interval and, for every $x\in X$, $E(x)$ is a \emph{simple} eigenvalue of
$H_\Lambda(x)$ (equivalently, $E(x)=E_{j(x)}^\Lambda(x)$ for some locally constant index $j(x)$). We call $X$
the \emph{phase interval} of the segment and $ \mathrm{im}(\Gamma):=E(X) $ its \emph{image}. We call the
segments in the next proposition {\em regular}; their normalized eigenfunctions have sub-exponentially small
endpoint values.
\end{definition}

Fix $0<\alpha_0<1/2$ for the rest of the paper. For each odd $N_0$, let $n_0=N_0^{\alpha_0}+O(1)$ be the least odd integer
not less than $N_0^{\alpha_0}$. 

\begin{proposition}\label{ls:opposite}
Let $p_3$ be such that $p_3>\max\{p_1,\alpha_0(p_2+2)+\tau_\omega\}$. For every sufficiently large odd $N_0=2\widetilde{N}_0+1$, the operator
$H_{\Lambda_{N_0}}$ has increasing and decreasing spectral segments $\Gamma_{\pm}$ with
common image
\begin{align*}
 I=(E_1-\varepsilon_0,E_1+\varepsilon_0),\qquad \varepsilon_0=N_0^{-(p_2+p_3)}.
\end{align*}
Their slopes have absolute value greater than $N_0^{-p_2}$, and their phase intervals have length at least
$2\varepsilon_0/M_V$. Suppose $(x,E)\in \Gamma_\pm$ and denote the associated normalized eigenfunction
by~$\psi$. Then
\begin{align}\label{ls:endpoint}
 |\psi(-\widetilde{N}_0)|+|\psi(\widetilde{N}_0)|\leq2e^{-n_0 L_*/2}.
\end{align}
Its extension by zero to $\ell^2(\Z)$ is a unit vector $\widetilde\psi$ with
\begin{align}\label{ls:quasimode}
 \|(H(x)-E)\widetilde\psi\|\leq2e^{-n_0L_*/2}.
\end{align}
\end{proposition}

It follows from Corollary~\ref{cor:genRZug} and periodicity that there are at least $(1/2-\delta)N_0$ segments crossing $E_1$ of each orientation. 
The difficulty is to ensure that the segment is non-resonant for the two edge intervals 
\begin{align}\label{def:I-I+}
 I_-=[-\widetilde{N}_0,-\widetilde{N}_0+n_0-1],\qquad I_+=[\widetilde{N}_0-n_0+1,\widetilde{N}_0].
\end{align}
We do this by first locating all the zeros of $F_{I_\pm}(\cdot,E_1)$ in the fixed strip $\T_{\varrho_0}$. There are at most $2\CZ n_0$ such zeros by Lemma~\ref{lem:count-bounds}. 
For each such zero $\zeta=x_{\zeta}+iy_{\zeta}$, we delete all the segments that cross $E_1$ in a short phase interval containing $x_{\zeta}$. The next lemma shows that there are not too many segments entering each such phase interval, and hence the total number deleted is at most $o(N_0)$. Denote by $x_*$ the crossing of a surviving segment at energy $E_1$. 
Then $F_{I_\pm}(x,E_1)$ are zero-free in the real neighborhood of $x_*$. 
Lemma \ref{lem:no0nonres} implies $I_{\pm}$ are non-resonant at any $(x,E)$ that is close enough to $(x_*,E_1)$.

To be precise, Lemma~\ref{ls:rank} estimates the number of eigenvalues in a small energy window at a fixed $x$, and then uses transversality of Rellich curves to convert the estimate into a bound on the number of crossings in a short phase interval.

\begin{lemma}\label{ls:rank}
Fix any constant $A_0>1$ and
\begin{align*}
 p_4>p_2+2+A_0\tau_\omega.
\end{align*}
For every $E\in\mathcal G$ and all sufficiently large odd $n$, depending on $E$, every interval $\Lambda\subset \Z$ with $4n\leq N=|\Lambda|\leq n^{A_0}$ and every
 $x\in \T$ satisfy
\begin{align}\label{ls:rankbound}
 \operatorname{rank}\mathbf1_{[E-n^{-p_4},\, E+n^{-p_4}]} (H_\Lambda(x))\leq (\CZ +2)n.
\end{align}
Moreover, if $E\in \mathcal{G}$ and an interval $X\subset\T$ satisfies
\begin{align}\label{ls:crossing-interval}
 M_V|X|<\min\{n^{-p_4} ,2N^{-p_1}\},
\end{align}
then there are at most $(\CZ +2)n$ crossings $E_j^\Lambda(x)=E$, $x\in X$, counting all  $j$.
\end{lemma}

\begin{proof}
Let $n=2\tilde{n}+1$. For a fixed phase $x$, let
\begin{align*}
 \mathcal R_x:=\{k\in\Lambda: \dist(E,\spec H_{\Lambda_n+k}(x))<2n^{-p_4}\}.
\end{align*}
For each $k\in\mathcal R_x$, Lemma~\ref{lem:transcom} shows that
$F_{\Lambda_n}(\cdot,E)$ has a real zero within distance $2n^{p_2-p_4}$ of $x+k\omega$ on~$\T$. For distinct $k,k'\in\Lambda$,
\begin{align*}
 \|(k-k')\omega\|_\T\geq c_\omega N^{-\tau_\omega} \geq c_\omega n^{-A_0\tau_\omega}\gg n^{p_2-p_4}.
\end{align*}
Hence the real zeros near different $x+k\omega$ are distinct.
The  upper bound on the number of zeros  from~\eqref{eq:globbud} implies 
\begin{align*}
 |\mathcal R_x|\leq \CZ n.
\end{align*}
Including the neighborhood of the endpoints of $\Lambda$, the number is still controlled:
\begin{align*}
\widetilde{\mathcal{R}}_x:=\mathcal R_x\cup \{k\in\Lambda:\dist(k,\partial\Lambda)<n\}
\end{align*}
satisfies $|\widetilde{\mathcal{R}}_x|\leq(\CZ +2)n$.

If $k\in\Lambda\setminus \widetilde{\mathcal{R}}_x$, then $\Lambda_n+k\subset\Lambda$, and $\Lambda_n+k$ is non-resonant at any $(x,E')$ with $|E'-E|\leq n^{-p_4}$.
In fact, if $\Lambda_n+k$ is resonant at $(x,E')$ for some $E'$, $|E'-E|\leq n^{-p_4}$, then Lemma~\ref{lem:res_close} implies
\begin{align*}
    \mathrm{dist}(E', \sigma(H_{\Lambda_n+k}(x)))\leq M_V\rho_n\ll n^{-p_4}.
\end{align*}
This would imply $\mathrm{dist}(E,\sigma(H_{\Lambda_n+k}(x)))\leq 2n^{-p_4}$, hence $k\in \mathcal{R}_x$, a contradiction.
Cramer's rule and the determinant upper bound now yield
\begin{align*}
 |G_{\Lambda_n+k}(x,E';k,k-\tilde{n})|+|G_{\Lambda_n+k}(x,E';k,k+\tilde{n})| \leq2\exp\bigl(Cn^a-L_*\tilde{n}\bigr).
\end{align*}
The Poisson formula~\eqref{eq:Poisson} then shows that every normalized eigenfunction $\psi$ of $H_\Lambda(x)$ with
eigenvalue $E'$ in $[E-n^{-p_4},E+n^{-p_4}]$ satisfies 
$|\psi(k)|\leq e^{-L_*n/3}$ for $k\in\Lambda\setminus \widetilde{\mathcal{R}}_x$. 
Let $\Pi$ be the spectral projection in \eqref{ls:rankbound}. Summing over an orthonormal eigenbasis implies
\begin{align*}
 \|\mathbf1_{\Lambda\setminus \widetilde{\mathcal{R}}_x}\Pi\|_{\rm HS}^2 \leq N\operatorname{rank}(\Pi)e^{-2L_*n/3} \leq N^2e^{-2L_*n/3}<1.
\end{align*}
In conclusion, the restriction map from $\operatorname{ran}\Pi$ to $\ell^2(\widetilde{\mathcal{R}}_x)$ is injective. 
Hence $\operatorname{rank}\Pi\leq|\widetilde{\mathcal{R}}_x|\leq (\CZ +2)n$.

To prove the second assertion, fix an interval $X\subset\T$.
Every Rellich function $E_j^{\Lambda}(\cdot)$ crossing $E$ in $X$ stays within $2N^{-p_1}$ of $E$ throughout $X$, by
\eqref{ls:crossing-interval} and $\sup_x |\partial_x E_j^{\Lambda}(x)|\leq M_V$. Since $E$ is not in the forbidden zone, $E_j^{\Lambda}(x)$ is strictly monotone in the energy strip $[E-2N^{-p_1}, E+2N^{-p_1}]$ by
Lemma~\ref{lem:transcom}. 
Distinct crossings are therefore associated with different indices. 
At any fixed $x_0\in X$, these curves give distinct eigenvalues in
\begin{align*}
 [E-M_V|X|,E+M_V|X|]\subset [E-n^{-p_4},E+n^{-p_4}], 
\end{align*}
and their number is bounded by \eqref{ls:rankbound}.
\end{proof}

We are now ready to prove the existence of regular segments. Recall $N_0=2\widetilde{N}_0+1$ and the auxiliary scale $n_0=N_0^{\alpha_0}+O(1)$.  
In the proof of the proposition, we shall exclude crossings near the zeros of the two endpoint intervals $I_{\pm}$, see \eqref{def:I-I+}, at energy~$E_1$. 

\begin{proof}[Proof of Proposition~\ref{ls:opposite}]
Set $\Lambda=\Lambda_{N_0}$. Let $C_{N_0}^+$ and $C_{N_0}^-$ denote the number of  positive, respectively,  negative $E_1$-crossings of $H_\Lambda$.  
By periodicity, transversality, and Corollary~\ref{cor:genRZug},
\begin{align}\label{eq:rzugsb}
 C_{N_0}^+=C_{N_0}^-\geq (1/2-\delta) N_0.
\end{align}
In Lemma~\ref{ls:rank} fix $E=E_1$, $n=n_0$, $N=N_0$ and $A_0=1/\alpha_0$. 
Choose $p_4$ such that
\begin{align*}
    p_3>\alpha_0p_4>\alpha_0(p_2+2)+\tau_{\omega}.
\end{align*}
Set $r_0=N_0^{-p_3}$ and $\varepsilon_0=N_0^{-p_2}r_0$.

By the global zero count \eqref{eq:globbud}, the two endpoint determinants $F_{I_\pm}(\cdot,E_1)$
have at most $2\CZ  n_0$ zeros in total in the fixed strip $\T_{\varrho_0}$, counted with multiplicity. For each such zero $\zeta=x_{\zeta}+iy_{\zeta}$, we remove all the segments $(x, E_j^{\Lambda}(x))$ which cross the energy $E_1$ at a phase that belongs to $[x_\zeta-4r_0,x_\zeta+4r_0]$. 
Each such phase interval has length $8r_0$. Since $p_3>\max\{\alpha_0p_4,p_1\}$, 
\begin{align}\label{ls:rchoice}
 8M_Vr_0<\min\{n_0^{-p_4},2N_0^{-p_1}\}.
\end{align}
Hence Lemma~\ref{ls:rank} implies it contains at most $(\CZ +2)n_0$ crossings of the segments. 
Hence the total number of segments removed is at most
\begin{align}\label{ls:deleted}
2\CZ (\CZ +2)n_0^2=o(N_0).
\end{align}
In view of \eqref{eq:rzugsb}, crossings of both signs remain. Choose one segment $(x,E_{j_\pm}^{\Lambda}(x))$ of each sign,  which crosses $E_1$ at phases $x_\pm$, respectively. 
By construction, both endpoint determinants $F_{I_-}(\cdot,E_1)$ and $F_{I_+}(\cdot,E_1)$ do not vanish in the intervals $[x_\pm-4r_0,x_\pm+4r_0]$.
Restrict the two segments to the smaller energy window $(E_1-\varepsilon_0, E_1+\varepsilon_0)$ and denote them by $\Gamma_{\pm}$.
Their inverse branches $x_{\pm}(\cdot)$ satisfy 
\begin{align*}
 \|x_\pm(E)-x_\pm\|_\T \leq N_0^{p_2}|E-E_1| \leq N_0^{p_2}\varepsilon_0 \leq r_0.
\end{align*}
Note $r_0\gg \widehat{\rho}_{n_0}$.
Lemma \ref{lem:no0nonres}, applied to $n=n_0$ and $F_{I_{\pm}}(\cdot,E_1)$, implies that throughout $(x,E)\in \Gamma_{\pm}$,
\begin{align*}
 \log|F_{I_\pm}(x,E)| \geq n_0L_*-C_3n_0^a\geq n_0L_*/2.
\end{align*}
For a normalized eigenfunction $\psi$ at $(x,E)\in \Gamma_{\pm}$, the Poisson formula at the left endpoint and Cramer's rule give
\begin{align*}
 |\psi(-\widetilde{N}_0)| 
 &\leq |G_{I_-}(x,E;-\widetilde{N}_0, -\widetilde{N}_0+n_0-1)| \\
 &=|F_{I_-}(x,E)|^{-1} \leq e^{-n_0L_*/2}.
\end{align*}
The right estimate is analogous.  This proves \eqref{ls:endpoint}.
The phase lengths of the two segments are clearly at least $2\varepsilon_0/M_V$. 
Finally, extending $\psi$ by zero to $\ell^2(\Z)$ leaves a residual $(H(x)-E)\widetilde\psi$ supported only at $-\widetilde N_0-1$ and $\widetilde N_0+1$. Thus
\begin{align*}
 \|(H(x)-E)\widetilde\psi\|^2 =|\psi(-\widetilde N_0)|^2 +|\psi(\widetilde N_0)|^2 \leq2e^{-n_0L_*},
\end{align*}
which proves \eqref{ls:quasimode}. 
\end{proof}

\section{Localization to the resonant region}
\label{sec:localization}

Following the procedure from~\cite{FSW90, GS11}, we will intersect the two regular segments from Proposition~\ref{ls:opposite} via a translation on~$\Z$ and thus produce a local maximum of a Rellich function on a larger lattice interval.  This section prepares some preliminary facts about localization in a finite volume. Throughout this section, $(x,E)\in\T\times \mathcal{J}$ is arbitrary but fixed, and the estimates are uniform in this pair. We use notation and constants from Section~\ref{sec:basictools}.

\begin{definition}
For odd $n=2\tilde n+1$, set
\begin{align}\label{eq:singular-delta}
 \delta_Z=\frac1{4(\CZ +1)},\qquad b_n=\lfloor(1/2-\delta_Z)n\rfloor.
\end{align}
We call a site $t\in\Z$ \emph{singular} at scale $n$ if every $\Lambda_n+k$ with $|k-t|\leq b_n$ is resonant at $(x,E)$.
\end{definition}

Throughout this
section, fix $A_1>1$, which controls the length of the ambient interval $\Lambda$.  
The next two lemmas hold for all sufficiently large odd $n$, with thresholds depending on $A_1$ and uniform in $(x,E)\in\T\times J$.

\begin{lemma}\label{lem:badtrans}
Let $\Lambda\subset \Z$ be an interval such that $|\Lambda|\leq 2n^{A_1}$. Among the intervals $\Lambda_n+k$ with
$k\in\Lambda$, at most $\CZ n$ are resonant.
\end{lemma}

\begin{proof}
By Lemma~\ref{lem:Cartan}, each resonant $\Lambda_n+k$ implies a zero of $F_{\Lambda_n}(\cdot,E)$ within $\rho_n$ of
$x+k\omega$. 
For distinct $k,k'\in\Lambda$, the Diophantine condition implies that 
\begin{align*}
 \|(k-k')\omega\|_\T \geq c_\omega(2n^{A_1})^{-\tau_\omega} \gg\rho_n.
\end{align*}
For sufficiently large $n$, we have $\rho_n<\varrho_0$. Hence, these zeros lie in $\T_{\varrho_0}$ and are distinct.
The global zero count upper bound for $F_{\Lambda_n}(\cdot, E)$ in \eqref{eq:globbud} implies the claimed bound on the number of $k$.
\end{proof}

The next lemma uses this count  to show that the singular sites in a polynomially large interval can be covered by at most $\CZ $ intervals of length $2n+1$.

\begin{lemma}\label{lem:regions}
Let $\Lambda$ be an interval of length $|\Lambda|=N$, $4n\leq N\leq n^{A_1}$. Its singular sites are contained
in at most $\CZ$ intervals
\begin{align}\label{eq:regioncov}
 \mathcal R=\bigcup_{i=1}^s[t_i-n,t_i+n], \qquad s\leq \CZ ,
\end{align}
whose centers $t_i$ are singular sites in $\Lambda$.
\end{lemma}

\begin{proof}
Suppose that there are $\CZ +1$ singular sites $t_1,\ldots,t_{\CZ +1}$ in $\Lambda$ with pairwise distances 
at least $n$. Since $2b_n<n$, the sets
\begin{align*}
 \{k\in\Z:|k-t_i|\leq b_n\}, \qquad 1\leq i\leq \CZ +1,
\end{align*}
are disjoint. By the definition of singular sites, all the corresponding intervals $\Lambda_n+k$ are resonant. 
These centers $k$ lie in $\Lambda+[-b_n,b_n]$, whose diameter is bounded by $2n^{A_1}$.
Lemma~\ref{lem:badtrans} therefore implies that 
\begin{align*}
 (\CZ +1)(2b_n+1)\leq \CZ n.
\end{align*}
This is impossible for large $n$, since
\begin{align*}
 (\CZ +1)(1-2\delta_Z)>\CZ .
\end{align*}
Thus any collection of singular sites with pairwise distances at least $n$ contains at most $\CZ $ points. 
Choose a maximal such collection $t_1,\ldots,t_s$. Every other singular site is within distance less than $n$
of one of them. This proves \eqref{eq:regioncov}.
\end{proof}

A non-resonant interval $I\subset \Z$ at $(x,E)$ leads to a {\em good} Green function $G_I(x,E)$ by Cramer's rule.  ``Good'' here refers to the bounds in~\eqref{eq:GreenI}, which are stable under perturbations in phase and energy by a standard Neumann series argument. We provide the details for the reader's convenience. Note that at a nonsingular site~$t$, some non-resonant interval $\Lambda_n+k$ satisfies $|t-k|\leq b_n$, so both endpoints are at distance at least $\delta_Z n/2$ from~$t$ for large~$n$. This then  leads to~\eqref{eq:endpointGreen} in that case. 

\begin{lemma}\label{lem:stableGreen}
Recall constant $C_3>0$ in Lemma~\ref{lem:Cartan}. For $n$ sufficiently large, if $I$ is non-resonant at $(x,E)$ and $n\leq|I|\leq2n$,
then for any real phase $x'\in\T$ and any real energy $E'$ satisfying
\begin{align}\label{eq:varxE}
 M_V\|x'-x\|_\T+|E'-E| \leq e^{-6C_3n^a},
\end{align}
the resolvent $G_I(x', E')$ exists and, for all $i,j\in I$,
\begin{align}\label{eq:GreenI}
 |G_I(x',E';i,j)| \leq \exp(6C_3n^a-L_*|i-j|/2).
\end{align} 
In particular, if $\Lambda_n+k$ is non-resonant at $(x,E)$ and $|t-k|\leq b_n$, then
\begin{align}\label{eq:endpointGreen}
 |G_{\Lambda_n+k}(x',E';t,k-\tilde n)| +|G_{\Lambda_n+k}(x',E';t,k+\tilde n)| 
 \leq e^{-\delta_Z L_* n/4}.
\end{align}
\end{lemma}

\begin{proof}
For simplicity, write $m=L_*/2$. 
For $i,j$ in $I$, Cramer's rule, \eqref{eq:wtoben}, and
\eqref{eq:fcgood} give
\begin{align*}
 |G_I(x,E;i,j)|\leq\exp(2C_3n^a-L_*|i-j|)
\end{align*} for sufficently large $n$.
Define the weighted matrix norm
\begin{align*}
 \|T\|_m=\max_i\sum_j|T(i,j)|e^{m|i-j|}.
\end{align*}
This norm is submultiplicative. Since $m<L_*$, the preceding estimate implies
\begin{align*}
 \|G_I(x,E)\|_m \leq 2n e^{2C_3n^a}\leq e^{4C_3n^a}.
\end{align*}
The perturbation
\begin{align*}
 W=H_I(x')-H_I(x)-(E'-E)\mathrm{Id}
\end{align*}
is diagonal and satisfies
\begin{align*}
 \|W\|_m \leq M_V\|x'-x\|_\T+|E'-E|.
\end{align*}
By \eqref{eq:varxE}, the Neumann series converges and implies
\begin{align*}
 \|G_I(x',E')\|_m \leq\frac{e^{4C_3n^a}}{1-e^{4C_3n^a}\|W\|_m} \leq e^{6C_3 n^a},
\end{align*}
which proves \eqref{eq:GreenI}. 
Finally, if $|t-k|\leq b_n$, then 
\begin{align*}
 \min\{|t-(k-\tilde n)|,\, |t-(k+\tilde n)|\} \geq\tilde n-b_n\geq\delta_Z n-\frac12.
\end{align*}
Thus,  \eqref{eq:GreenI} implies
\begin{align*}
 &|G_{\Lambda_n+k}(x',E';t,k-\tilde n)| +|G_{\Lambda_n+k}(x',E';t,k+\tilde n)| \\
 &\qquad\leq 2\exp\left(6C_3n^a-m\left(\delta_Z n-\frac12\right)\right) \\
 &\qquad\leq \exp\left(-\delta_Z L_* n/4\right),
\end{align*}
for sufficiently large $n$.
\end{proof}

The following proposition quantifies how most of the mass of an eigenfunction is located near the resonant sites.

\begin{proposition}\label{prop:localization}
Let $\Lambda=[u,v]$ be an interval of length $N$, with $4n\leq N\leq n^{A_1}$, and let
the nonempty interval 
$\Lambda_{\rm res}\subset\Lambda$ contain every singular site of $\Lambda$ at scale $n$ at $(x,E)$.
Suppose furthermore that the two endpoint intervals
\begin{align*}
 I_-=[u,u+n-1],\qquad I_+=[v-n+1,v]
\end{align*}
are non-resonant at $(x,E)$. 
For every real phase $x'\in\T$ and real energy $E'$
satisfying \eqref{eq:varxE}, every $\ell^2$-normalized eigenfunction $H_\Lambda(x')\psi=E'\psi$
satisfies
\begin{align}\label{eq:tail}
 \|\mathbf1_{\Lambda\setminus\Lambda_{\rm res}}\psi\| \leq \sqrt N\,e^{-\delta_Z L_*n/4},
 \qquad \|\mathbf1_{\Lambda_{\rm res}}\psi\|^2\geq\tfrac{3}{4},
\end{align}
for all sufficiently large $n$. Moreover,
\begin{align}\label{eq:spatialdec}
 |\psi(t)| \leq e^{-\delta_Z L_* \dist(t,\Lambda_{\rm res})/8}, \qquad t\in\Lambda\setminus\Lambda_{\rm res}.
\end{align}
The constants
and thresholds are uniform in $(x,E)\in \T\times \mathcal{J}$.
\end{proposition}

\begin{proof}
Let $t\in\Lambda\setminus\Lambda_{\rm res}$ be a  nonsingular site.

First, suppose $t$ is not close to either endpoint that $\min\{t-u,v-t\}\geq\tilde n+b_n$. 
By nonsingularity, there exists a non-resonant interval
$\Lambda_n+k\subset\Lambda$ with $|t-k|\leq b_n$. 
The Poisson formula and the Green's function estimates in \eqref{eq:endpointGreen} yield
\begin{align*}
 |\psi(t)|\leq e^{-\delta_Z L_*n/4} \max\left(|\psi(k-\tilde{n}-1)|, |\psi(k+\tilde{n}+1)|\right).
\end{align*}
If $u\leq t<u+\tilde n+b_n$, we instead use the non-resonant interval $I_-$. Since $\psi(u-1)=0$, the Poisson
formula reduces to
\begin{align*}
 \psi(t) =G_{I_-}(x',E';t,u+n-1)\psi(u+n).
\end{align*}
Since $t\leq u+\tilde n+b_n-1$ \begin{align*}
 u+n-1-t\geq\tilde n-b_n+1\geq\delta_Z n.
\end{align*}
The Green's function estimate \eqref{eq:GreenI} implies
\[
|\psi(t)|\leq e^{-\delta_ZL_*n/4}|\psi(u+n)|. 
\]
The right endpoint is treated
similarly using $I_+$.

In all three cases, the boundary values in the Poisson formula are at distance at most $2n$ from $t$.
Consequently,
\begin{align}
 |\psi(t)| \leq e^{-\delta_ZL_*n/4} \max_{\substack{s\in\Lambda\\|s-t|\leq2n}}|\psi(s)|,
 \qquad t\in\Lambda\setminus\Lambda_{\rm res}.\label{eq:1stepdec}
\end{align}
Since $\|\psi\|_\infty\leq1$, this proves the first inequality in \eqref{eq:tail}. The second follows from
\begin{align*}
 \|\mathbf{1}_{\Lambda_{\rm res}}\psi\|^2 \geq 1-Ne^{-\delta_ZL_*n/2} \geq\frac{3}{4}
\end{align*}
for sufficiently large $n$, since $N\leq n^{A_1}$.
To prove \eqref{eq:spatialdec}, define
\begin{align*}
 S(R)=\sup\{|\psi(t)|:t\in\Lambda,\ \dist(t,\Lambda_{\rm res})>R\},
\end{align*}
with the convention that the supremum is zero when the set is empty.
By \eqref{eq:1stepdec},
\begin{align}\label{eq:S<S}
 S(0)\leq e^{-\delta_ZL_*n/4}, \qquad S(R+2n)\leq e^{-\delta_ZL_*n/4}S(R).
\end{align}
Indeed, if $\dist(t,\Lambda_{\rm res})>R+2n$, then every $s$ with $|s-t|\leq2n$ satisfies
$\dist(s,\Lambda_{\rm res})>R$.
Iterating \eqref{eq:S<S} implies
\begin{align*}
 S(2nk)\leq e^{-\delta_ZL_*n(k+1)/4}, \qquad k\geq0.
\end{align*}
For $t\notin\Lambda_{\mathrm{res}}$, let $d=\dist(t,\Lambda_{\mathrm{res}})$ and choose $k=\lceil d/(2n)\rceil-1$. Then
$2nk<d\leq2n(k+1)$, and hence
\begin{align*}
 |\psi(t)|\leq S(2nk) \leq e^{-\delta_ZL_*n(k+1)/4} \leq e^{-\delta_ZL_*d/8},
\end{align*}
as claimed. 
\end{proof}

\section{A local maximum with localized eigenfunctions}
\label{sec:local-max}

We intersect the two regular segments from Proposition~\ref{ls:opposite}  by translating one of them by a multiple of~$\omega$. 
By means of the corresponding eigenfunctions we then generate local maxima of Rellich functions at a larger scale. 
The analysis in Sec.~\ref{sec:localization} helps us to choose the larger scale properly which guarantees the local maximum is isolated from the other Dirichlet eigenvalues, and the corresponding eigenfunctions have exponentially small boundary values.
This mechanism goes back to second-order perturbation theory, cf.~\cite[Fig.~2a and b]{FSW90}, and was developed
nonperturbatively in~\cite{GS11}. 

\subsection{Intersecting two regular spectral segments}
\

We first make the two opposite segments intersect by means of a spatial translation.

\begin{lemma}\label{lem:fobmat}
Let $\lambda_\pm:X_\pm\to I$ be two spectral-segment eigenvalue branches with common energy image $I$ satisfying
\begin{align*}
 0<\lambda_+'\leq M_V,\qquad -M_V\leq\lambda_-'<0.
\end{align*}
Under \eqref{eq:DC}, for every integer $m_0\geq1$ there exists an integer $m$ such that
\begin{align*}
 m_0\leq m<m_0+C_{\omega,M_V}|I|^{-\tau_\omega},
\end{align*}
an energy $E_m$ in the middle half of $I$, and a phase $x$ such that
\begin{align}\label{eq:matchCX}
 \lambda_+(x)=\lambda_-(x+m\omega)=E_m.
\end{align}
The phases $x$ and $x+m\omega$ have distance at least $|I|/(4M_V)$ from the endpoints of their respective
segments.
\end{lemma}

\begin{proof}
Choose real inverse branches $x_\pm(E')$ of the Rellich segments and define $h(E'):=x_-(E')-x_+(E')$. Then
\begin{align*}
 h'(E')=\frac1{\lambda_-'(x_-(E'))} -\frac1{\lambda_+'(x_+(E'))}\leq-\frac2{M_V}.
\end{align*}
Let $I_{\rm mid}$ denote the middle half of $I$. We have
\begin{align*}
 |h(I_{\rm mid})|\geq\frac{|I|}{M_V}.
\end{align*}
We claim that for every integer $m_0$, for every interval $(y-\varepsilon/2,y+\varepsilon/2)\mathrel{\subset}\T$ of length $0<\varepsilon\leq1$, there exists $0\leq \ell\leq C_{\omega}\varepsilon^{-\tau_{\omega}}$ such that 
$\|y-(m_0+\ell)\omega\|_{\T}<\varepsilon/2$, in particular $(m_0+\ell)\omega$ enters the interval.

In fact, let $\{p_j/q_j\}$ be the continued fraction convergents of $\omega$. Take the smallest $j$ such that $q_j>4\varepsilon^{-1}$.
Since $\{\ell p_j/q_j\}_{\ell=0}^{q_j-1}=\{k/q_j\}_{k=0}^{q_j-1}$,
there exists $\ell$, $0\leq \ell\leq q_j-1$, such that
\[\|y-m_0\omega-\ell \tfrac{p_j}{q_j}\|_{\T}\leq \frac{1}{2q_j}.\]
By properties of continued fractions, we also have
\[\|\ell\omega-\ell \tfrac{p_j}{q_j}\|_{\T}=\ell |\omega-\tfrac{p_j}{q_j}|<\frac{1}{q_{j+1}}.\]
This implies $\|y-(m_0+\ell)\omega\|_{\T}<2/q_j<\varepsilon/2$. Finally note that the choice of $j$ implies $q_{j-1}\leq 4\varepsilon^{-1}$ and by the Diophantine condition
\[\ell<q_j\leq c_{\omega}^{-1}q_{j-1}^{\tau_{\omega}}\leq C_{\omega} \varepsilon^{-\tau_{\omega}},\]
as claimed.

Finally applying the claim to a sub-interval of $h(I_{\rm mid})$ with length $\varepsilon=|I|/(2M_V)$ implies the existence of some $0\leq \ell\leq C_{\omega,M_V}|I|^{-\tau_{\omega}}$ such that $m\omega=(m_0+\ell)\omega =h(E_m) \in h(I_{\rm mid}) \pmod 1$. Taking $x=x_+(E_m)$ implies that 
\eqref{eq:matchCX}. The last assertion follows from $|\lambda_\pm'|\leq M_V$ and $I_{\rm mid}$ is at least $|I|/4$ away from the endpoints of $I$.
\end{proof}

The next lemma turns this type of crossing of translated Rellich functions with opposite slopes of one scale into a local maximum of
a Rellich function at a larger scale. The presence of several competing ordered Rellich functions makes the extraction of that local maximum delicate. \cite[Lemma~15.4]{GS11} deals with this difficulty via two separate pigeonhole arguments at both endpoints of the phase interval.    The following lemma simplifies that construction by means of a single pigeonhole argument at equally spaced energy levels.

\begin{lemma}\label{lem:pollocmax}
Let $\lambda_1(x)<\cdots<\lambda_Q(x)$ be $C^1$ functions on $X=[x_0-h,x_0+h]\subset\T$, with $|\lambda_j'|\leq M_V$. 
 Suppose  $a_+,a_-\in C^1(X)$ satisfy
\begin{align*}
 a_+(x_0)=a_-(x_0)=E_0,\qquad \varepsilon\leq a_+'(x)\leq M_V,\qquad -M_V\leq a_-'(x)\leq-\varepsilon, \qquad 0<\varepsilon\leq M_V,
\end{align*}
and 
\begin{align*}
 \min_j|\lambda_j(x)-a_\pm(x)|\leq\eta \leq\frac{h\varepsilon}{4(Q+1)}
\end{align*}
for all  $x\in X$. 
Then there are $x_-<x_0<x_+$ in $X$  such that for some~$j$
\begin{align}
\label{eq:lambda var}
 \max_{[x_-,x_+]}\lambda_j \geq\max\{\lambda_j(x_-),\lambda_j(x_+)\} +\frac{h\varepsilon}{2(Q+1)}.
\end{align}
In particular, every maximum lies in the interior of $[x_-,x_+]$ and 
$|\lambda_j(x_0)-E_0|\leq 3M_Vh$.
\end{lemma}

\begin{proof}
Set
\begin{align*}
\Delta=\frac{h\varepsilon}{Q+1},\qquad E_k=E_0-k\Delta,\qquad 1\leq k\leq Q+1.
\end{align*}
For each $E_k$, there exist $x_k^-<x_0<x_k^+$ in $X$ such
that
\begin{align*}
 a_+(x_k^-)=a_-(x_k^+)=E_k.
\end{align*}
As $E_k$ increases, the corresponding phase interval shrinks. Hence, if $k<\ell$, then
$E_k>E_\ell$ and
\begin{align*}
 x_\ell^-<x_k^-<x_0<x_k^+<x_\ell^+.
\end{align*}
At each  $E_k$, choose  $i_k^-,i_k^+$ satisfying
\begin{align*}
 |\lambda_{i_k^-}(x_k^-)-E_k|\leq \eta,\qquad |\lambda_{i_k^+}(x_k^+)-E_k|\leq \eta,
\end{align*}
and define the {\em admissible index} as 
\begin{align*}
 j_k:=\min\{i_k^-,i_k^+\}.
\end{align*}
By the ordering of the eigenvalues,
\begin{align*}
\lambda_{j_k}(x_k^-),\lambda_{j_k}(x_k^+) \leq E_k+\eta,
\end{align*}
while at least one of these values is at least 
$E_k-\eta$. Thus
\begin{align*}
E_k-\eta\leq \max\{\lambda_{j_k}(x_k^-),\lambda_{j_k}(x_k^+)\} \leq E_k+\eta.
\end{align*}
There are $Q+1$ energy levels but only $Q$ possible admissible indices. So $j_k=j_\ell=:j$ for some $k<\ell$. Let
\begin{align*}
 x_-=x_\ell^-,\qquad x_+=x_\ell^+.
\end{align*}
The lower level $E_\ell$ bounds $\lambda_j$ from above at both outer points $x_-,x_+$, while the higher level $E_k$ ensures that $\lambda_j$ reaches at least $E_k-\eta$ at one of the inner points $x_k^-,x_k^+$. Consequently,
\begin{align*}
 \max_{[x_-,x_+]}\lambda_j -\max\{\lambda_j(x_-),\lambda_j(x_+)\} \geq E_k-E_\ell-2\eta \geq \Delta-2\eta \geq \Delta/2.
\end{align*}
Thus every maximizing point lies in the interior of $[x_-,x_+]$.
Finally, since $j=j_\ell$, at least one of $x_-,x_+$ satisfies $|\lambda_j(x_\pm)-E_\ell|\leq\eta$. Therefore,
\begin{align*}
 |\lambda_j(x_0)-E_0| \leq M_Vh+|E_\ell-E_0|+\eta \leq M_Vh+h\varepsilon+\eta \leq 3M_Vh.
\end{align*}
This completes the proof.
\end{proof}

\subsection{A local maximum with localized eigenfunctions}
\

We now place the two translated spectral segments in a common  interval $\Lambda$ with Dirichlet boundary conditions.

\begin{prop}\label{prop:loc_max}
Let $E_1\in\mathcal G$ be as in
Section~\ref{sec:p_by_c}.
Recall the constants $\delta_Z$ in \eqref{eq:singular-delta}, $L_*$ in \eqref{eq:positive-LE}, and $C_5$ in 
Proposition~\ref{prop:FSWgap}.
Let constant $c:=\delta_ZL_*/32$, and integer $D\geq 3$ be such that
\begin{align}\label{def:constant_D}
D>2c^{-1}(C_5+1).
\end{align}
Then, for every sufficiently large odd
$N_0=2\widetilde N_0+1$, there exist intervals
$0\in\Lambda_{\rm res}\subset\Lambda$, $|\Lambda_{\rm res}|=q$ and $|\Lambda|=Q$, satisfying
\begin{align*}
 q\geq N_0,\qquad Dq\leq Q\leq2Dq,
\end{align*}
a phase $x_*\in\T$, and an eigenvalue branch
$\lambda(x)=E_j^\Lambda(x)$ such that

\begin{enumerate}
\item $E_*:=\lambda(x_*)$ is a local max of $\lambda(\cdot)$, in particular $\lambda'(x_*)=0$.
\item $E_*$ is away from the other eigenvalues:
\begin{align}\label{eq:g_gap}
 \dist(E_*,\spec H_\Lambda(x_*)\setminus\{E_*\})\geq e^{-C_5 q}=:g.
\end{align}
\item $|E_*-E_1|\to0$ as $N_0\to\infty$.
\item With $r=e^{-(C_5+1)q}$, the normalized eigenfunctions corresponding to $\lambda(x)$ satisfy
\begin{align}\label{eq:psi_Lam_boundary}
 \sup_{|x-x_*|\leq r}\bigl(|\psi(x,\min\Lambda)|+|\psi(x,\max\Lambda)|\bigr)\leq 2e^{-cQ}=:\epsilon_\psi.
\end{align}

\item Let $\widetilde\psi(x)$ denote the extension of $\psi(x)$ by zero to $\ell^2(\Z)$.
Let $\widetilde{M}_V:=\|V''\|_\infty+4\|V'\|_\infty^2$.
Then, for $|x-x_*|\leq r$,
\begin{align}\label{eq:H-E_psi_small}
 \|(H(x)-E_*)\widetilde\psi(x)\|\leq \frac{\widetilde{M}_V r^2}{g}+\epsilon_\psi.
\end{align}
\end{enumerate}
\end{prop}

\begin{proof} 
The heuristic ideas are as follows:
we select two spatial scales by pigeonholing. At the smallest scale $n_0\mathrel{\asymp}N_0^{\alpha_0}$, the singular sites are covered by
at most $\CZ $ intervals of radius $n_0$. Among $\CZ +1$ disjoint annuli between successive scales $R_j$ and
$R_{j+1}\asymp4DR_j$, one contains no centers of these intervals. We choose $\Lambda_{\rm res}$ at the smaller
scale and $\Lambda$ at the larger scale. Then $Dq\leq Q\leq2Dq$, with a nonsingular region of length comparable to
$Q$ between $\Lambda_{\rm res}$ and each endpoint. Nonsingularity near the outer edges permits a small
adjustment of the endpoints so that their length-$n_0$ intervals are non-resonant.

We first choose constant $A_1$ as in
Section~\ref{sec:localization}, such that
\begin{align}\label{eq:choose_A1}
 A_1>\alpha_0^{-1}\max\{1,\tau_\omega(p_2+p_3)\}.
\end{align}
Apply Proposition~\ref{ls:opposite} at the initial length $N_0=2\widetilde N_0+1$. It implies two regular
spectral segments, denoted by $(x,E_{\pm}(x))$, with common image $(E_1-\varepsilon_0,E_1+\varepsilon_0)$,  $\varepsilon_0=N_0^{-(p_2+p_3)}$, slopes of opposite signs and absolute
value at least $N_0^{-p_2}$, and phase intervals of length at least $2\varepsilon_0/M_V$. By \eqref{ls:quasimode}, the zero extensions of their eigenfunctions, denoted by $\widetilde{\psi}_{\pm}(x)$, have small residual
\begin{align}\label{eq:ini_residual}
 \|(H(x)-E_{\pm}(x))\widetilde{\psi}_{\pm}(x)\|\leq 2e^{-n_0L_*/2}=:\eta_0.
\end{align}
Apply Lemma~\ref{lem:fobmat} with $m_0=2N_0$ and
$\lambda_\pm=E_\pm$ to make these two segments intersect.
Let $m$ and $x_m$ be the resulting translation and phase, and let
\begin{align*}
 a_+(x)=E_+(x),\qquad a_-(x)=E_-(x+m\omega).
\end{align*}
These branches correspond to the intervals
\begin{align*}
 \Lambda_0=[-\widetilde N_0,\widetilde N_0], \qquad \Lambda'_0=m+\Lambda_0.
\end{align*}
At the matching phase,
\begin{align*}
 a_+(x_m)=a_-(x_m)=E_m,
 \qquad |E_m-E_1|\leq\varepsilon_0/2.
\end{align*}
Both branches $a_\pm$ are defined on
the common phase interval
\begin{align}\label{def:Xm}
 X_m=
 \left[
 x_m-\frac{\varepsilon_0}{4M_V},
 x_m+\frac{\varepsilon_0}{4M_V}
 \right],
\end{align}
on which
\begin{align*}
 N_0^{-p_2}\leq a_+'\leq M_V,\qquad
 -M_V\leq a_-'\leq-N_0^{-p_2}.
\end{align*}
Moreover, Lemma\ref{lem:fobmat} with $m_0=2N_0$ yields
\[
2N_0\leq m<2N_0+ C_{\omega,M_V}\varepsilon_0^{-\tau_\omega}.
\]
Thus the two initial intervals $\Lambda_0$ and $\Lambda'_0$ are disjoint and lie within polynomial distance of the origin.
We now define the scales
\begin{align*}
R_0=2m, \qquad R_j=\lceil (4D)^j R_0 \rceil, \qquad 1\leq j\leq \CZ +1.
\end{align*}
All these scales are bounded by a fixed multiple of $m$, and hence by a fixed power of $N_0$.
By the choice of $A_1$ in \eqref{eq:choose_A1} and Lemma~\ref{lem:fobmat} one has for large $N_0$,
\begin{align*}
 6R_{\CZ +1}+1\leq n_0^{A_1}.
\end{align*}
Apply Lemma~\ref{lem:regions} at $(x_m,E_m)$, with  $n=n_0$, on $[-3R_{\CZ +1},3R_{\CZ +1}]$. 
Since $E_m\in \mathcal{J}$ for large $N_0$, all singular sites are covered by intervals
\begin{align*}
 \bigcup_i[t_i-n_0,t_i+n_0],
\end{align*}
with at most $\CZ $ centers $t_i$.
Consider the following $\CZ +1$ disjoint annuli
\begin{align*}
 \{t\in\Z: 2R_j\leq|t|<2R_{j+1}\}, \qquad 0\leq j\leq \CZ .
\end{align*}
By pigeonholing, one of them contains no centers of any singular intervals. 
Fix such an index $j$ and define
\begin{align*}
 \Lambda_{\rm res}=[-3R_j,3R_j], \qquad q:=6R_j+1.
\end{align*}
Since $n_0\ll R_j$, every singular interval whose center satisfies $|t_i|<2R_j$ is contained in
$\Lambda_{\rm res}$, while those with $|t_i|\geq 2R_{j+1}$ are disjoint from $[-R_{j+1},R_{j+1}]$.
Consequently, every site in
\begin{align}\label{eq:nonsingular_buffer}
 [-R_{j+1},R_{j+1}]\setminus\Lambda_{\rm res}
\end{align}
is nonsingular at scale $n_0$ for $(x_m,E_m)$. 
Our choice of $R_0$ also ensures that both initial intervals $\Lambda_0,\Lambda'_0$ lie in $\Lambda_{\rm res}$.

Next, we choose the Dirichlet endpoints using the nonsingularity of the region \eqref{eq:nonsingular_buffer}. In
particular, the sites $-R_{j+1}+n_0$ and $R_{j+1}-n_0$ are nonsingular. 
Thus each admits a nearby non-resonant interval of length $n_0$. Let $u$ be the left endpoint of the first interval and $v$ the right endpoint of the second one. 
Since $n_0=2\widetilde n_0+1$ and $b_{n_0}\leq\widetilde n_0$, we have
\begin{align*}
 -R_{j+1}\leq u\leq-R_{j+1}+n_0,\qquad R_{j+1}-n_0\leq v\leq R_{j+1}.
\end{align*}
Consequently,
\begin{align*}
 I_-=[u,u+n_0-1],\qquad I_+=[v-n_0+1,v]
\end{align*}
are non-resonant at $(x_m,E_m)$.
Set $\Lambda=[u,v]$ and $Q:=|\Lambda|$. Then $\Lambda_{\rm res}$ contains all singular sites of $\Lambda$, and
\begin{align}\label{eq:Q_length}
 Dq\leq Q\leq2Dq,\qquad \dist(\partial\Lambda,\Lambda_{\rm res})\geq Q/4.
\end{align}
Indeed, $Q=2R_{j+1}+O(n_0)$, so $Q/q\to4D/3$ as $N_0\to\infty$.
Also $\dist(\partial\Lambda,\Lambda_{\rm res})\geq (1-3/(4D))R_{j+1}+O(n_0)\geq Q/4$, since $4D\geq12$ and $n_0/R_j\to 0$.

By Proposition~\ref{prop:localization},  any normalized eigenfunction $\psi$ of $H_{\Lambda}(x)\psi=E'\psi$, whenever $(x,E')$ is a small perturbation of $(x_m,E_m)$, is localized within $\Lambda_{\rm res}$.
In fact, recall that $a=1-\theta/2<1$. Recall constant $6C_3$ is from \eqref{eq:varxE}, and set
\begin{align*}
 h_0=e^{-6C_3n_0^a},\qquad \widetilde h_0=\frac{h_0}{10M_V}.
\end{align*}
For every real phase $x$ and real energy $E'$ with
\begin{align}\label{eq:xmEm_perturb}
 \|x-x_m\|_\T\leq\frac{h_0}{2M_V},\qquad |E'-E_m|\leq\frac{h_0}{2},
\end{align}
the left side of \eqref{eq:varxE} is at most $h_0\leq e^{-6C_3n_0^a}$. 
By Proposition~\ref{prop:localization}, we have 
\begin{align}\label{eq:concLres}
    \|\mathbf{1}_{\Lambda_{\rm res}}\psi\|^2\geq 3/4,
\end{align}
and the exponential decay~\eqref{eq:spatialdec}.

For sufficiently large $N_0$, we have $\widetilde{h}_0\ll \varepsilon_0$, so both branches $a_\pm$ are defined on
$\widetilde{X}_m:=[x_m-\widetilde h_0,x_m+\widetilde h_0]\subset X_m$, see \eqref{def:Xm}. 
Extend their normalized eigenfunctions by zero to $\Lambda$. 
Let $T$ be the right shift on $\ell^2(\Z)$: $(Tu)(j)=u(j-1)$, then by covariance, $H(x)T^k=T^k H(x+k\omega)$ for any $k\in \Z$. 
By \eqref{eq:ini_residual}, for any $x\in \widetilde{X}_m$,
\begin{align*}
\dist(a_+(x),\spec H_\Lambda(x))\leq \|(H_{\Lambda}(x)-E_+(x))\widetilde{\psi}_+(x)\|\leq \eta_0,
\end{align*}
and 
\begin{align*}
\dist(a_-(x),\spec H_\Lambda(x))\leq \|(H_{\Lambda}(x)-E_-(x+m\omega))(T^m \widetilde{\psi}_-(x+m\omega))\|\leq \eta_0,
\end{align*}
hence 
\begin{align*}
 \dist(a_\pm(x),\spec H_\Lambda(x))\leq\eta_0.
\end{align*}
Moreover, since $a<1$, for large $N_0$, 
\begin{align*}
 \eta_0\leq\frac{\widetilde h_0N_0^{-p_2}}{4(Q+1)}.
\end{align*}
Applying Lemma~\ref{lem:pollocmax} with lower slope bound $N_0^{-p_2}$, we obtain an eigenvalue
$\lambda=E_j^\Lambda$ attaining a maximum $E_*=\lambda(x_*)$ on an interval
\begin{align*}
 [x_-,x_+]\subset \widetilde{X}_m,
\end{align*}
with $x_*$ in its interior and
\begin{align*}
 |\lambda(x_m)-E_m|\leq3M_V\widetilde h_0.
\end{align*}
Since $|x_*-x_m|\leq\widetilde h_0$, one has 
\begin{align}\label{eq:E*-Em}
 |E_*-E_m| \leq |\lambda(x_*)-\lambda(x_m)|+|\lambda(x_m)-E_m|\leq  4M_V\widetilde h_0=2h_0/5,
\end{align}
for large $N_0$.
Hence $(x_*,E_*)$ satisfies \eqref{eq:xmEm_perturb}. 
Suppose there is another eigenvalue $E_{j'}^{\Lambda}(x_*)$ such that $(x_*, E_{j'}^{\Lambda}(x_*))$ also satisfies \eqref{eq:xmEm_perturb}. 
Then both their normalized eigenfunctions would be concentrated on $\Lambda_{\rm res}$ in the sense of \eqref{eq:concLres}. 
Proposition~\ref{prop:FSWgap} would imply
\[|E_*-E_{j'}^{\Lambda}(x_*)|\geq e^{-C_5q}.\]
Therefore for sufficiently large $N_0$, 
\begin{align}\label{eq:E*_gap}
 \dist(E_*,\spec H_\Lambda(x_*)\setminus\{E_*\}) \geq \min\left\{\frac{h_0}{10}, e^{-C_5q}\right\}=e^{-C_5q}=g.
\end{align}
The last the equality above follows from $q\geq N_0\gg n_0^a$.

Recall $r=e^{-(C_5+1)q}\ll \min(g,h_0)$ for large $N_0$.
For $|x-x_*|\leq r$, $|\lambda(x)-E_*|\leq M_Vr$. Consequently, by \eqref{eq:E*-Em},
\begin{align*}
 |x-x_m|&\leq r+\widetilde h_0\leq h_0/(2M_V),\\
 |\lambda(x)-E_m|&\leq M_Vr+2h_0/5<h_0/2.
\end{align*}
Thus $(x,\lambda(x))$ satisfies \eqref{eq:xmEm_perturb}
throughout $|x-x_*|\leq r$.
The exponential decay estimate \eqref{eq:spatialdec},
together with \eqref{eq:Q_length}, then yields
\eqref{eq:psi_Lam_boundary}.
For (3), note that
\[|E_*-E_1|\leq |E_*-E_m|+|E_m-E_1|\leq 2h_0/5+\varepsilon_0/2\to 0,\]
as $N_0\to\infty$. 

Since
\begin{align*}
\|H_\Lambda(x)-H_\Lambda(x_*)\|
\leq M_Vr,\qquad |x-x_*|\leq r,
\end{align*}
the spectral gap $g$ at $x_*$, see \eqref{eq:E*_gap}, remains at least
$g-2M_Vr\geq g/2$ throughout this interval.

The second derivative formula for a simple eigenvalue,
see~\cite[eq.~(2.11)]{FSW90}, yields
\begin{align*}
|\lambda''(x)|
\leq \|V''\|_\infty
+\frac{4\|V'\|_\infty^2}{g}
\leq \frac{\widetilde{M}_V}{g},
\qquad |x-x_*|\leq r.
\end{align*}
Since $\lambda'(x_*)=0$, we obtain
\begin{align}\label{eq:lambda-E*_r2/g}
|\lambda(x)-E_*| \leq \frac{\widetilde{M}_Vr^2}{g},\qquad |x-x_*|\leq r.
\end{align}
Extending the corresponding $\psi(x)$
by zero to $\ell^2(\Z)$, we have by \eqref{eq:psi_Lam_boundary} and \eqref{eq:lambda-E*_r2/g} that
\begin{align*}
\|(H(x)-E_*)\widetilde\psi(x)\|
\leq |\lambda(x)-E_*|+\epsilon_{\psi}\leq \frac{\widetilde{M}_V r^2}{g}+\epsilon_{\psi}.
\end{align*}
This proves \eqref{eq:H-E_psi_small}.
\end{proof}

\section{The DOS mass near a maximum}
\label{sec:dos-contradiction}

Extending the eigenfunctions from Proposition~\ref{prop:loc_max} by zero produces compactly supported 
quasimodes parametrized by phases in some small interval. We first recall the well-known idea that a family of such quasimodes leads to a lower bound on the local DOS mass. 

\begin{lemma}\label{lem:dcnonf}
Let $X\subset\T$  and $\Lambda\subset\Z$ with $|\Lambda|=Q$. Suppose $E_*\in \R$ and there exists constant $\epsilon>0$ such that for every $x\in X$ there is a unit vector $\psi(x)$ supported in $\Lambda$ such that
\begin{align*}
 \|(H(x)-E_*)\psi(x)\|\leq\epsilon.
\end{align*}
Then
\begin{align}\label{eq:doskompMass}
 \nu_\omega([E_*-2\epsilon,E_*+2\epsilon]) \geq\frac{3|X|}{4Q}.
\end{align}
\end{lemma}

\begin{proof}
Let $\Pi(x)$ be the spectral projection of $H(x)$ onto $[E_*-2\epsilon,E_*+2\epsilon]$. 
Since
\begin{align*}
 \|(H(x)-E_*)\psi(x)\|^2 \geq 4\epsilon^2\|(\mathrm{Id}-\Pi(x))\psi(x)\|^2,
\end{align*}
the spectral theorem and $\|\psi(x)\|=1$ yield
\begin{align*}
 \langle \Pi(x)\psi(x),\psi(x)\rangle\geq\frac {3}{4},
\end{align*}
for $x\in X$. Since $\psi(x)$ is supported in $\Lambda$,
\begin{align*}
 \frac{3}{4}\mathbf{1}_X(x) \leq\sum_{j\in\Lambda}\langle \Pi(x)\delta_j,\delta_j\rangle.
\end{align*}
Integrating and using covariance implies that 
\begin{align*}
 \frac{3|X|}{4} \leq\sum_{j\in\Lambda}\int_\T \langle \Pi(x)\delta_j,\delta_j\rangle\,dx 
 =Q\cdot \nu_\omega([E_*-2\epsilon,E_*+2\epsilon])
\end{align*}
as claimed. 
\end{proof}

The final contradiction follows from Lemma~\ref{lem:dcnonf}.
At a local maximum of a sufficiently isolated eigenvalue,
the variation of the eigenvalue is at most quadratic in the phase.
If the corresponding eigenfunctions have small boundary values,
this forces too much DOS mass into a small energy interval.

\begin{proof}[Proof of Theorem~\ref{thm:cantor}]\
Recall that we assumed $J_0$ is an interval of constant acceleration in $\Sigma_{\omega}\cap \mathcal{O}_+$, and $E_1$ is a generalized eigenvalue in $\mathrm{int}(J_0)$.
By Proposition \ref{prop:loc_max}, there exists normalized $\widetilde{\psi}(x)$ supported in $\Lambda$ with $|\Lambda|=Q$, such that
\begin{align*}
\|(H(x)-E_*)\widetilde{\psi}(x)\|
\leq \frac{\widetilde{M}_V r^2}{g}+\epsilon_\psi=:\epsilon, \text{ for all } |x-x_*|\leq r.
\end{align*}
Since $E_*\to E_1\in\mathrm{int}(J_0)$ and $\epsilon\to 0$,
the interval $[E_*-2\epsilon,E_*+2\epsilon]$ is contained
in $J_0$ for large $N_0$.

\noindent Lemma~\ref{lem:dcnonf} applied to
$X=[x_*-r,x_*+r]$, and \eqref{eq:doslip} applied to $J_0$, yield
\begin{align*}
\frac{3r}{2Q}
\leq\nu_\omega([E_*-2\epsilon,E_*+2\epsilon])
\leq 4C_{J_0}\epsilon
\leq 4C_{J_0}\left(\frac{\widetilde{M}_V r^2}{g}+\epsilon_\psi\right).
\end{align*}
Consequently,
\begin{align*}
\frac{1}{q}\lesssim \frac{1}{Q}
\leq 4C_{J_0}\left(\widetilde{M}_V\frac{r}{g}+\frac{\epsilon_\psi}{r}\right)
\lesssim \left(
e^{-q}
+e^{-(cD-(C_5+1))q}
\right).
\end{align*}
This is a contradiction as $N_0\to\infty$ since $D>2c^{-1}(C_5+1)$.
\end{proof}

\end{document}